\documentclass[a4paper,11pt]{article}
\usepackage{jcappub}
\usepackage{amsmath}
\usepackage{amssymb}
\usepackage[T1]{fontenc}
\usepackage{siunitx}
\usepackage{listings}
\usepackage[normalem]{ulem}
\usepackage{bm}
\usepackage{cases}
\usepackage{upgreek}
\usepackage{multirow}
\usepackage{enumerate}
\usepackage{lineno}

\newcommand{\Nsh}{N_{\text{sh}}}
\newcommand{\vrms}{v_{\text{rms}}}
\newcommand{\Pgw}{\mathcal{P}_{\text{gw}}}

\newcommand{\bk}{\bm{k}}
\newcommand{\bp}{\bm{p}}
\newcommand{\bq}{\bm{q}}

\newcommand{\tv}{\tilde{v}}

\newcommand{\detav}{\Delta\eta_{\text{v}}}

\title{Signatures of the speed of sound on the gravitational wave power spectrum from sound waves}

\author[a]{Lorenzo Giombi,}
\author[a]{Jani Dahl,}
\author[a,b]{Mark Hindmarsh}

\affiliation[a]{Department of Physics and Helsinki Institute of Physics,\\PL 64, FI-00014 University of Helsinki, Finland}
\affiliation[b]{Department of Physics and Astronomy, University of Sussex,\\Brighton BN1 9QH, United Kingdom}

\emailAdd{lorenzo.giombi@helsinki.fi}
\emailAdd{jani.dahl@helsinki.fi}
\emailAdd{mark.hindmarsh@helsinki.fi}

\abstract{
Future space-based interferometers offer an unprecedented opportunity to detect signals from the stochastic gravitational wave background originating from a first-order phase transition at the electroweak scale. The phase transition is accompanied by a change of the equation of state from that of pure radiation. In this work we study the effect of this change on the power spectrum of gravitational waves generated by the sound waves in the plasma during the acoustic phase of the transition. We carry out an analytic calculation assuming that the sound speed and the fluid shear-stress that sources tensor perturbations remain approximately constant during the acoustic phase. The effect of a softer equation of state is twofold:
(i) a scale-independent suppression of the power spectrum at all scales, due to the modified propagation of both sound and gravitational waves and 
(ii) the peak of the spectrum moves to smaller frequencies as the equation of state becomes softer. 
The power-law indices of the spectrum at small and large scales are unaffected by the softening of the equation of state. Our work improves the current estimation of the gravitational waves power spectrum from first order phase transitions and expands the possible scenarios of transitions that can be tested by gravitational wave detectors.
}

\begin{document}
\maketitle
\flushbottom

\section{Introduction}\label{sec:introduction}
Relic gravitational waves from the early Universe  propagate without scattering throughout the Universe from the moment they are generated until today, offering a picture of the Universe before recombination and of the physical mechanism that produced them~\cite{Maggiore:1999vm, Caprini:2019egz}.  
The direct detection of relic gravitational waves is an ambitious challenge for modern gravitational wave astronomy, and a compelling science objectives of current and future missions searching beyond the bounds set by LIGO~\cite{LIGOScientific:2007fwp, Moore:2014lga, Caprini:2024ofd}. 
The recent NANOGrav experiments on pulsar timing arrays reported, for the first time in 2023, evidence for a signal from the stochastic gravitational wave background (SGWB) at $\si{\nano\Hz}$ frequencies~\cite{EPTA:2023fyk, EPTA:2023xxk, NANOGrav:2023gor, NANOGrav:2023hvm}.
In the next decade, the space-based gravitational wave detector Laser Interferometer Space Antenna (LISA)~\cite{LISA:2017pwj} will be launched and operative to observe signals at the $\si{\milli\Hz}$ frequencies. 
Besides receiving many signals from different astrophysical sources, 
LISA will be listening for possible signals in the SGWB generated at the electroweak scale~\cite{LISACosmologyWorkingGroup:2022jok, Caprini:2015zlo, Caprini:2019egz}. One of the sources of gravitational waves at these scales is a cosmological first order phase transition (FOPT), which is the focus of this article.

Cosmological FOPTs~\cite{Kirzhnits:1972iw, Kirzhnits:1972ut, Kirzhnits:1976ts, Kibble:1980mv} are violent phenomena in the early Universe 
that can drive, among others, the processes of electroweak baryogenesis~\cite{Cohen:1990it, Cline:2018fuq, Cline:2006ts} 
and primordial black holes formation~\cite{Kodama:1982sf, Hawking:1982ga, Khlopov:1998nm, Franciolini:2021nvv, Liu:2021svg, Lewicki:2023ioy},  
leaving characteristic imprints on the SGWB~\cite{Hindmarsh:2020hop}. 
The dynamics of a FOPT is characterized by the nucleation, expansion, and eventual merging of bubbles of the new (low-temperature) phase~\cite{Witten:1980ez, Guth:1981uk, Steinhardt:1980wx}.
During the expansion, the potential energy difference between the two phases  
is converted into kinetic and thermal energy of the fluid around bubbles, 
which thereby becomes a source of anisotropic stress~\cite{Steinhardt:1981ct}, and thus gravitational radiation~\cite{Witten:1984rs, Hogan:1986qda}. Gravitational waves are produced through three different processes:
(i) collision of bubbles~\cite{Kosowsky:1991ua, Kosowsky:1992rz, Kosowsky:1992vn, Huber:2008hg, Jinno:2017fby, Cutting:2018tjt}, (ii) sound waves~\cite{Hindmarsh:2013xza,Hindmarsh:2015qta,Hindmarsh:2016lnk,Hindmarsh:2017gnf,Hindmarsh:2019phv, Jinno:2020eqg, Jinno:2022mie}, and 
(iii) vortical turbulence~\cite{Kosowsky:2001xp, Gogoberidze:2007an, Caprini:2009fx, Caprini:2009yp, RoperPol:2019wvy, Auclair:2022jod, Cutting:2019zws}. 
Due to the long-lasting nature of the acoustic phase, sound waves are thought to be the dominant source of gravitational waves~\cite{Weir:2017wfa}. This is also confirmed by numerical simulations of non-relativistic flows, which indicate that the expansion of bubbles generates only compressional modes~\cite{Hindmarsh:2013xza, Hindmarsh:2015qta, Hindmarsh:2017gnf}.

The universal shape of the power spectrum of gravitational waves produced by sound waves can be understood with few physical considerations~\cite{RoperPol:2019wvy, Caprini:2009fx,Auclair:2022jod, RoperPol:2022iel, Hindmarsh:2016lnk,Hindmarsh:2019phv,Sharma:2023mao}. The mean bubble spacing $R_*$ inherently defines the characteristic wavelength of sound waves and the frequency scale $k_{p} = 2\pi R_*^{-1}$ where most of the energy is located; the power spectrum has therefore a peak amplitude at $k_\star \sim k_p$.
Well below the peak frequency, at periods greater than the lifetime of the sound wave source, causality enforces $\Pgw\propto k^3$.
Above the peak, the formation of shocks in the fluid imposes the universal scaling $\Pgw\propto k^{-3}$.  For periods between the source lifetime and the source characteristic wavelength, a scaling argument shows that $\Pgw\propto k$ \cite{Sharma:2023mao,RoperPol:2023dzg}.
Semi-analytic modelling also indicates a growing peak with a $k^9$ power-law on the low-frequency side \cite{Hindmarsh:2016lnk,Hindmarsh:2019phv,Sharma:2023mao,RoperPol:2023dzg}

The semi-analytic model that has been most successful in understanding and reproducing the features of the
acoustic gravitational waves power spectrum is the Sound Shell Model (SSM)~\cite{Hindmarsh:2016lnk}.
The key observation of the model is the fact that the fluid dragged by the bubbles' interfaces forms
shells of compression and rarefaction that keep propagating by inertia even
after the collision with other bubbles, when the phase boundaries 
disappear and the pressure gradients driving the expansion vanish.
Long-lasting sound waves in the plasma emerge as a linear superposition 
of many sound shells, whose initial conditions are set at the time of the bubbles' boundaries collision.
Non-linearities in the fluid~\cite{Cutting:2019zws, Gogoberidze:2007an, Caprini:2009yp, RoperPol:2019wvy}
become important on a timescale $\tau_{\text{nl}} = R_*/v_{\text{rms}}$, 
with $v_{\text{rms}}$ the enthalpy-weighted root mean square (RMS) fluid velocity, and are not considered in this work.

The predictions of the SSM on the gravitational wave power spectrum~\cite{Hindmarsh:2019phv} 
are in good agreement with numerical simulations of weak phase transitions in a flat static Universe~\cite{Hindmarsh:2013xza, Hindmarsh:2015qta}. For long-lasting sources, the expansion of the Universe becomes increasingly important with the source duration, and it has only been included in the simulations as a conformal re-scaling of the energy-momentum tensor of the cosmic fluid~\cite{Hindmarsh:2013xza, Hindmarsh:2015qta, Hindmarsh:2017gnf, Espinosa:2010hh}. 
Semi-analytical extensions of the SSM instead, already managed to include the effects of the Universe expansion on the propagation of gravitational waves at leading order in $R_*\mathcal{H}_*$, the fractional bubble radius over the size of the causal Universe at the time of the transition~\cite{Cai:2023guc, RoperPol:2023dzg}. This introduces 
a friction contribution 
on the propagation of gravitational waves that suppresses  the tensor power 
spectrum homogeneously at all scales~\cite{RoperPol:2023dzg,Sharma:2023mao,Guo:2020grp}.

In a fluid dominated by ultra-relativistic particles, the damping of gravitational wave due to the Hubble friction is the only contribution to be expected, at leading order in $R_*\mathcal{H}_*$, from the expansion of the Universe. However, in a real scenario, the Universe might deviate from a conformal equation of state: the symmetry-breaking process that drives the phase transition introduces new degrees of freedom in the cosmic fluid and modifies particle interactions inside the plasma~\cite{Hindmarsh:2020hop}; this can lead to a substantial change in the energy density and pressure of the cosmic fluid that drives the Universe expansion~\cite{Ares:2020lbt}. When the equation of state (EOS) relating pressure and energy density deviates significantly from the ultra-relativistic case, other contributions to the gravitational wave power spectrum become important. 

In this work we study the implications of a softer equation of state on the power spectrum of acoustically generated gravitational waves in a cosmological phase transition. We consider the scenario where the typical size of the bubbles comprising the low-temperature phase is well within the causal horizon at the time of the transition $R_*\mathcal{H}_* \ll 1$. Given the large difference in scales, we treat perturbatively the dynamic equations for sound and gravitational waves in the short-wavelength parameter $R_*\mathcal{H}_*$. To pursue an analytic calculation, we further simplify the evolution of the EOS, assuming that it becomes instantaneously softer at the beginning of the acoustic phase $\eta_*$, and that it remains constant for a finite time until $\eta_{\text{r}}$, when it abruptly transitions to the pure radiation case. 

At leading order in $R_*\mathcal{H}_*\ll 1$, the softening of the equation of state affects the energy density of gravitational waves in two different ways: (i) it increases the expansion rate of the Universe 
and thereby decreases the total energy of the Universe at $\eta_*$ compared to the critical density today; (ii)  it introduces a friction in the propagation of sound waves that damps the source of the shear stress over time. 
Contributions beyond the linear order become relevant when the bubbles grow to sizes $R_*\mathcal{H}_*\lesssim \mathcal{O}(1)$, and they include, in addition, the effects of scalar induced tensor perturbations~\cite{Baumann:2007zm}. We dedicate this article to the analysis of leading order contributions, and postpone the discussion of next-to-leading order effects to a future work.

Throughout this paper we use units $c = \hbar = 1$. We further adopt the 
mostly positive signature for the metric $(-, +,+,+)$. 
Greek letters $(\mu, \nu, \dots)$ will be used for four-dimensional 
tensorial indices, while latin letters $(i, j, \dots)$ for three-dimensional indices.
Finally, we find convenient to use conformal time $\eta$ as time coordinate, with $a(\eta)d\eta = dt$ 
and $a(\eta)$ the scale factor in a Friedmann-Lema\^{i}tre-Robertson-Walker Universe.
We use a prime to denote derivative with respect to $\eta$, e.g. $a^\prime = da/d\eta$.


\section{Gravitational wave power spectrum from sound waves}\label{sec:GW_PS}
We consider tensor perturbations $h_{ij}$ propagating thought an expanding Friedmann-Lema\^{i}tre-Robertson-Walker (FLRW) Universe~\cite{Durrer:2004fx, Brandenberger:2003vk}
\begin{equation}
    ds^2 = a^2(\eta)\left[ - d\eta^2 + \left(\delta_{ij} + h_{ij}\right) dx^i dx^j\right].
\end{equation}
During the acoustic phase, the main source of gravitational waves is provided by the shear motion of the fluid~\cite{Hindmarsh:2019phv}. We assume that the matter in the Universe prior and during the transition can be described by a perfect fluid with energy-momentum tensor
\begin{equation}
    {T^\mu}_\nu = (e+p)u^\mu u_\nu + p{\delta^\mu}_\nu
\end{equation}
with 4-velocity $u^\mu$ and a barotropic equation of state (EOS) relating the pressure $p$ to the energy density $e$ as $p(e) = \omega e$, being $\omega$ a constant EOS parameter. 
A FOPT generates shear stress, transverse and traceless (TT) components of the energy-momentum tensor, and thereby sources gravitational waves through the linearized Einstein equation
\begin{equation}\label{eq:GW}
    \left(\partial_\eta^2 + 2\mathcal{H}\partial_\eta  - \nabla^2\right) h_{ij} = 16 \pi G a^2 \Lambda_{ij, \ell m}T_{\ell m},
\end{equation}
with $G$ the gravitational constant and $\mathcal{H} \equiv a^\prime/a$ the Hubble parameter in conformal time. The projection operator
\begin{equation}
    \Lambda(\hat{\bm{n}})_{ij, \ell m} = P_{i\ell}P_{jm} - \frac{1}{2}P_{ij}P_{\ell m},
\end{equation}
with $P_{ij}= \delta_{ij} - \hat{\bm{n}}_i\hat{\bm{n}}_j$ and $\hat{\bm{n}}_i$ the direction of propagation of the gravitational wave, projects the energy--momentum tensor onto the TT subspace~\cite{Maggiore:2007ulw}.

The intensity of the signal from a FOPT in the SGWB is determined by the energy density of gravitational waves produced during the transition, which we compute with the Isaacson formula~\cite{Isaacson:1968hbi, Isaacson:1968zza}
\begin{equation}
    e_{\text{gw}} = \frac{1}{32 \pi G a^2}\langle h^{\prime\,*}_{ij} h^\prime_{ij}\rangle.
\end{equation}
We assume that the mechanism of generation of gravitational waves from sound waves emerges as a stochastic process after the superposition of fluid motion from every bubble, and that the process is statistically homogeneous and isotropic. The symmetries of the system constrain the two-point correlation function of $h^\prime$ to the expression
\begin{equation}\label{eq:h_correlator}
    \langle \tilde{h}_{ij}^{\prime \, *} (\bm{k}_1, \eta) \tilde{h}_{ij}^{\prime}(\bm{k}_2, \eta)\rangle = (2\pi)^3  \delta(\bm{k}_1 - \bm{k}_2) P_{\tilde{h}^{\prime}}(\bm{k}_1, \eta),
\end{equation}
where $\tilde{h}_{ij} (\bm{k}, \eta)$ denotes the Fourier transform of the gravitational wave field
\begin{equation}
    \Tilde{h}_{ij}(\bm{k},\eta) = \int d^3\bm{x} \, h_{ij}(\bm{x},\eta) e^{-i\bm{k} \cdot \bm{x}} .
\end{equation}
Finally we define the fractional gravitational wave energy density $\Omega_{\text{gw}}\equiv e_{\text{gw}}/\Bar{e}$, with $\Bar{e}$ the critical energy density, and the power spectrum
\begin{equation}
    \Pgw \equiv \frac{d\Omega_{\text{gw}}}{d\ln k} =  \frac{1}{12\mathcal{H}^2}\frac{k^3}{2\pi^2}P_{\tilde{h}^\prime}.
\end{equation}

Analytic solutions to the wave equation~\eqref{eq:GW} depend on the expansion rate of the Universe. We simplify the equation by scaling out the Hubble friction term with the field redefinition $\ell_{ij} \equiv a(\eta) h_{ij}$~\cite{RoperPol:2023dzg} and seek solutions for the scaled tensor perturbations 
\begin{equation}\label{eq:GW_rescaled}
    \left(\partial_\eta^2  + k^2 -\frac{a^{\prime\prime}}{a}\right) \tilde{\ell}_{ij}(\bm{k}, \eta) = 16\pi G a^3\bar{w} \, \tilde{\mathcal{S}}_{ij}(\bm{k}, \eta),
\end{equation}
where we defined the dimensionless anisotropic stress $\tilde{\mathcal{S}}_{ij} \equiv  \Lambda_{ij,\ell m}T_{\ell m}/\Bar{w}$, with $w = e+p$ the enthalpy of the fluid. A bar is used to denote quantities evaluated on the FLRW background.  
We assume that the anisotropic stress turns on at an initial time $\eta_*$ and remains stationary for a finite time until $\eta_{\text{end}}$, when it rapidly switches off. 
The solution to the wave equation~\eqref{eq:GW_rescaled} can be then split as
\begin{equation}
  \tilde{\ell}_{ij}(\bm{k}, \eta) = 16 \pi G
    \begin{cases}
      {\displaystyle\int_{\eta_*}^\eta} d{\eta}_1\, a^3(\eta_1)\bar{w}(\eta_1)\, \tilde{\mathcal{S}}_{ij}(\bm{k}, \eta_1) G_k(\eta, \eta_1), & \eta_* <\eta <\eta_{\text{end}},\\
      &\\
      {\displaystyle\int_{\eta_*}^{\eta_{\text{end}}}} d\eta_1\, a^3(\eta_1)\bar{w}(\eta_1)\,\tilde{\mathcal{S}}_{ij}(\bm{k}, \eta_1) G_k(\eta, \eta_1), & \eta > \eta_{\text{end}}.
    \end{cases}       
\end{equation}
The Green's function $G_k(\eta, \eta_1)$  depends on the time evolution of the scale factor 
$a^{\prime\prime}/a$; its analytic solution is discussed in more details in Appendix~\ref{app:Green_function}. For a barotropic fluid with EOS $p=\omega e$, the scale factor evolves in time as $a(\eta) = a(\eta_*) (\eta/\eta_*)^{2/(1+3\omega)}$, and the Green's function has the closed expression
\begin{equation}\label{eq:Green}
    G_k(\eta, \eta_1) = \eta \, \dfrac{j_{\nu}(k\eta)y_{\nu}(k\eta_1) - j_{\nu}(k\eta_1)y_{\nu} (k\eta)}{\left[ \eta_1 j_{\nu}(k\eta_1)\right]^\prime y_{\nu}(k\eta_1) - \left[ \eta_1 y_{\nu}(k\eta_1)\right]^\prime j_{\nu}(k\eta_1)} \, \Theta(\eta -\eta_1), \qquad \nu =  \frac{1-3\omega}{1+3\omega},
\end{equation}
with $j_\nu$ and $y_\nu$ spherical Bessel functions of the first and second kind, and $\Theta(\eta -\eta_1)$ the Heaviside step function; the derivatives in the denominator are taken with respect to $\eta_1$.
The parameter $\nu \in [0, 1)$ describes the softening of the EOS and measures the deviation from the case of pure radiation $\omega = 1/3$, where the Universe expands conformally with $a^{\prime\prime} = 0$.

Once the source switches off at $\eta_{\text{end}}$, gravitational waves propagate freely in the Universe. At much later time $\eta \gg \eta_{\text{end}}$, neglecting contributions of order $1/(k\eta)$, the two-point correlation function can be written as~\footnote{The characteristic wavelength of perturbations that were inside the horizon at the time of the transition is much smaller than the Hubble radius at much later times, $k\eta\gg 1$.}
\begin{eqnarray}\label{eq:ll}
    \langle \tilde{h}^{*\, \prime}_{ij}(\eta, \bm{k}_1) \tilde{h}^{\prime}_{ij}(\eta, \bm{k}_2)  \rangle &=& \frac{1}{a^2(\eta)}\langle \tilde{\ell}^\prime_{ij}(\eta, \bm{k}_1) \tilde{\ell}^{\prime}_{ij}(\eta, \bm{k}_2)  \rangle \\
    &=& \frac{(16\pi G)^2}{a^2(\eta)}  \int_{\eta_*}^{\eta_{\text{end}}} d\eta_1   \int_{\eta_*}^{\eta_{\text{end}}} d\eta_2 \, G_{k_1}^\prime(\eta, \eta_1)G_{k_2}^\prime(\eta, \eta_2) \times \nonumber\\
    &&\qquad\times  a^3(\eta_1) a^3(\eta_2)\Bar{w}(\eta_1) \Bar{w}(\eta_2)  \left\langle\tilde{\mathcal{S}}^*_{ij}(\eta_1, \bm{k}_1) \tilde{\mathcal{S}}^{ij}(\eta_2, \bm{k}_2)\right\rangle,\quad
\end{eqnarray}
where the unequal time correlation (UETC) function of the shear stress, subject to the statistical isotropy and homogeneity that characterizes the system of sound waves, is
\begin{equation}\label{eq:U_s}
    \left\langle \tilde{\mathcal{S}}^*_{ij}(\eta_1, \bk) \tilde{\mathcal{S}}^{ij}(\eta_2, \bk^\prime)\right\rangle = U_{\mathcal{S}}(k, \eta_1, \eta_2) (2\pi)^3 \delta(\bk - \bk^\prime).
\end{equation}
Finally, we make explicit the time dependence of background quantities using the Friedmann equation $3\mathcal{H}^2 = 8\pi G e a^2$ and the  solutions 
\begin{equation}
    a(\eta) = a_* \left(\frac{\eta}{\eta_*} \right)^{1+\nu} , \qquad \bar{w}(\eta) = \bar{w}_* \left(\frac{a}{a_*} \right)^{-2\frac{2+\nu}{1+\nu}} ,
\end{equation}
with $a_* \equiv a(\eta_*)$ and $\bar{w}_* \equiv \bar{w}(\eta_*)$. Therefore, at times $\eta \gg \eta_{\text{end}}$, the dimensionless power spectrum of gravitational wave per logarithmic wavenumber can be written as
\begin{eqnarray}\label{P_hp}
    \Pgw &=&  3(1+\omega)^2\mathcal{H}_*^2 \left(\frac{a_*}{a}\right)^{4}\frac{\bar{e}_*}{\bar{e}} \frac{k^3}{2\pi^2} \iint_{\eta_*}^{\eta_{\text{end}}} d\eta_1  d\eta_2 \left(\frac{\eta_*^2}{\eta_1\eta_2}\right)^{1-\nu} \times \qquad\qquad\nonumber\\
    && \qquad\qquad\qquad\qquad\qquad\qquad\qquad \times\; G_{k}^\prime(\eta, \eta_1)G_{k}^\prime(\eta, \eta_2) U_{\mathcal{S}}(k, \eta_1, \eta_2).
\end{eqnarray}
After the sound waves dissipate at time $\eta_{\text{end}}$, gravitational waves start propagating freely with no source. However the Universe can take longer to transition toward the standard radiation dominated era, and the EOS describing the cosmic fluid can still be soft at time $\eta >\eta_{\text{end}}$. In a real transition, the change in the EOS happens smoothly but, for the purpose of an analytic estimation, we make the approximation that the EOS changes instantaneously at the radiation time $\eta_{\text{r}} >\eta_{\text{end}}$. 
We then consider the EOS parameter as a time-dependent distribution
\begin{equation}
  \omega(\eta) =
    \begin{cases}
      \omega & \eta_* < \eta < \eta_{\text{r}},\\
      1/3, & \eta_{\text{r}} < \eta,
    \end{cases}       
\end{equation}
while we remain agnostic about the value of the EOS parameter $\omega$ in the symmetric phase before the acoustic phase. 
With this approximation, at time $\eta > \eta_{\text{r}}$, we evaluate
\begin{equation}\label{eq:a_ar}
    \left(\frac{a_*}{a}\right)^{4}\frac{\bar{e}_*}{\bar{e}} =  \left(\frac{a_*}{a_{\text{r}}}\right)^{\frac{2\nu}{1+\nu}} .
\end{equation}
This factor represents the dilution of the background energy density $\Bar{e}$ due to the Universe expansion. The softer the EOS, the faster the expansion and the smaller $\Bar{e}_*$ compared to the critical density $\Bar{e}$.

\subsection{The UETC of the anisotropic stress}
We assume that the flow induced by sound waves in the plasma proceeds at  non-relativistic speed $\vert \bm{v}\vert \ll 1$, with $v^i$ the fluid 3-velocity. Numerical simulations of non-relativistic flows indicate that longitudinal modes are the most energetic, while transverse modes provide a negligible contribution to energy spectrum of the source~\cite{Hindmarsh:2013xza, Hindmarsh:2015qta, Hindmarsh:2017gnf}. We therefore restrict our analysis to the compressional components of the fluid velocity.
Upon projection onto TT space, only the transverse and traceless terms of the energy--momentum tensor can contribute to source gravitational waves.
Therefore, in momentum space, we can consider
\begin{equation}
    \tilde{\mathcal{S}}_{ij} (\eta, \bm{k}) = 
     \Lambda_{\ell m, i j} \int \frac{d^3 \bm{p}}{(2\pi)^3} \hat{p}_\ell \,\hat{q}_m  \tilde{v}(\eta, \bm{p}) \tilde{v}(\eta, -\bm{q}) 
\end{equation}
with $\bm{q} = \bm{p} - \bm{k}$ and $\tilde{v}(\eta,\bm{p}) = i \tilde{v}^i(\eta,\bm{p}) p_i/p \equiv i \hat{p}_i \tilde{v}^i(\eta,\bm{p})$, and the shear stress UETC
\begin{eqnarray}
    \langle \tilde{\mathcal{S}}^*_{ij}(\bm{k}, \eta_1) \tilde{\mathcal{S}}^{ij}(\bm{k}^\prime, \eta_2)\rangle &=&  \Lambda_{ij, k\ell}(\bm{k})  \Lambda_{ij,mn}(\bm{k}^\prime) \times\nonumber\\
    &&\times \int \frac{d^3 \bm{p_1}}{(2\pi)^3} \int \frac{d^3 \bm{p_2}}{(2\pi)^3} \hat{p}_1^k \hat{q}_1^\ell \hat{p}_2^m \hat{q}_2^n  \langle \tilde{v}^*_{1,\bm{p_1}} \tilde{v}^*_{1,-\bm{q_1}} \tilde{v}_{2,\bm{p_2}} \tilde{v}_{2,-\bm{q_2}} \rangle \label{pipi}
\end{eqnarray}
where $\bm{q_1} = \bm{p_1} - \bm{k}$ and $\bm{q_2} = \bm{p_2} - \bm{k}^\prime$; for shortness, we introduced the notation $\tilde{v}_{i,\bm{p_i}} \equiv \tilde{v}(\eta_i,\bm{p_i})$. 
 If we assume that velocity and density perturbations in the fluid are described by Gaussian statistics, we can use the Wick's theorem to rewrite the four-point correlation functions as a linear combination of two-point correlators:
\begin{equation}\label{eqs:GG}
    \langle \tilde{v}_{1,\bp}  \tilde{v}^*_{2,\bq} \rangle = C_{\tilde{v}\tilde{v}}(p, \eta_1, \eta_2)(2\pi)^3 \delta^3(\bp - \bq).
\end{equation}
The UETC of velocity perturbations $C_{\tilde{v}\tilde{v}}(p, \eta_1, \eta_2)$ depends on the initial conditions in the fluid, and will be discussed in more details in Section~\eqref{sec:correlators}. Finally
\begin{eqnarray}
    \langle \tilde{v}^*_{1,\bm{p_1}} \tilde{v}^*_{1,\bm{q_1}} \tilde{v}_{2,\bm{p_2}} \tilde{v}_{2,\bm{q_2}} \rangle &=& (2\pi)^6 \delta^3(\bk-\bk^\prime) \left[ \delta^3(\bm{p_1} -\bm{p_2}) + \delta^3(\bm{p_2} -\bm{q_1})\right] \times \nonumber\\
    && \qquad\qquad\qquad\qquad \qquad\qquad  \times \;C_{\tilde{v}\tilde{v}}(p_1, \eta_1, \eta_2)C_{\tilde{v}\tilde{v}}(q_1, \eta_1, \eta_2).\quad
\end{eqnarray}
The first Dirac delta function aligns the momenta $\bk^\prime$ and $\bk$, so that $\Lambda_{ij, k\ell}(\bm{k})  \Lambda_{ij,mn}(\bm{k}) = \Lambda_{k\ell, mn}(\bm{k})$. The delta functions inside the square brackets instead align the integration variables and allow the integration over $\bm{p_2}$ in equation~\eqref{eq:U_s} to be performed analytically, resulting in
\begin{equation} 
    U_{\mathcal{S}}(k, \eta_1, \eta_2) =  \int \frac{d^3 \bp}{(2\pi)^3}  (1-\mu_p^2) (1-\mu_q^2)  C_{\tilde{v}\tilde{v}}(p, \eta_1, \eta_2)C_{\tilde{v}\tilde{v}}(q, \eta_1, \eta_2),
    \end{equation}
with $\mu_p\equiv \hat{\bp}\cdot\hat{\bk}$ and $\mu_q\equiv \hat{\bm{q}}\cdot\hat{\bk}$. The integration over the azimuthal angle $\varphi$ is trivial, while the integration over the polar angle $\mu = \cos\theta$ can be changed to $q$ considering that $\mu_p = (p^2+k^2-q^2)/2pk$ and $1 - \mu_q^2 = (1 - \mu_p^2)p^2/q^2$. This way 
\begin{equation}\label{eq:U}
    U_{\mathcal{S}}(k, \eta_1, \eta_2) = \frac{1}{4\pi^2 k} \int_0^\infty dp \int_{\vert p -k\vert}^{p+k}dq (1-\mu_p^2)^2 \frac{p^3}{q}   C_{\tilde{v}\tilde{v}}(p, \eta_1, \eta_2)C_{\tilde{v}\tilde{v}}(q, \eta_1, \eta_2).
\end{equation}
The shear stress UETC is now completely determined by the UETC of velocity perturbations, whose evaluation is the topic of the next section.

\subsection{Shear stress correlators from sound waves}\label{sec:correlators}
Sound waves in the fluid are longitudinal perturbations in the velocity field $v^i$ and energy density field $e$. It is convenient to normalize the energy density with the average enthalpy density in the dimensionless variable 
\begin{equation}
    \lambda \equiv \frac{e(\bm{x}, \eta) - \bar{e}(\eta)}{\bar{w}(\eta)}.
\end{equation}
Conservation of energy and momentum $\nabla_\mu T^{\mu\nu} = 0$ provides the equations of motion for the fluid perturbation variables 
\begin{subequations}\label{eq:eom}
    \begin{eqnarray}
        \tilde{\lambda}^\prime_{\bm{p}} +  p\tilde{v}_{\bm{p}}  &=& 0, \label{eq:d_diff}\\
        \tilde{v}^\prime_{\bm{p}} - c_s^2p\tilde{\lambda}_{\bm{p}} + \frac{2\nu}{\eta}\tilde{v}_{\bm{p}}  &=& 0.\label{eq:v_diff}
    \end{eqnarray}
\end{subequations}
Equation~\eqref{eq:d_diff} and the first two terms in equation~\eqref{eq:v_diff} describe the propagation of stationary sound waves in a flat static background~\cite{Hindmarsh:2016lnk,  Hindmarsh:2019phv, Hindmarsh:2020hop, RoperPol:2023dzg}. 
The last term in equation~\eqref{eq:v_diff} is a general relativistic correction that represents the dissipation due to the expansion of the Universe~\footnote{In the exotic scenario where the EOS gets instead stiffer during the transition, the Universe decelerates and the velocities are amplified in time.}. 
This dissipation vanishes in a radiation dominated Universe, when $\nu =0$. 
As the EOS deviates that of pure radiation and gets softer, the scale factor develops a positive $a''/a$ which suppresses the peculiar motion of the fluid.  

When the bubbles comprising the new stable phase are well within the Hubble horizon at the time of the transition, we can consider $R_*\mathcal{H}_*\ll 1$. The dissipation of sound waves is the only general relativistic effect at the leading order in $R_*\mathcal{H}_*$. Beyond the leading order other effects become relevant, as the contribution from curvature perturbations~\cite{Durrer:2004fx}.
We postpone the analysis of these contributions to future studies. 

The system~\eqref{eq:eom} has analytic solution
\begin{subequations}\label{eq:sol_fluid}
    \begin{eqnarray}
    \Tilde{v}_{\bm{p}} &=& \left(\frac{\eta}{\eta_*}\right)^{-\nu} p\eta \Big[j_\nu(c_s p\eta) c_1 + y_\nu (c_s p\eta) c_2 \Big],\\
    \Tilde{\lambda}_{\bm{p}} &=& \frac{1}{c_s} \left(\frac{\eta}{\eta_*}\right)^{-\nu} p\eta \Big[j_{\nu-1} (c_s p\eta) c_1 + y_{\nu-1} (c_s p\eta) c_2 \Big], 
\end{eqnarray}
\end{subequations}
with $c_1$, $c_2$ real constants. Since we assumed $R_*\mathcal{H}_*\ll 1$, we can expand the spherical Bessel functions for large arguments (see Appendix~\ref{app:Green_function}). At leading order in $R_*\mathcal{H}_*$, the fluid perturbations~\eqref{eq:sol_fluid} can be expressed as a superposition of damped plane waves 
\begin{subequations}\label{eq:sound_waves}
\begin{eqnarray}
    \tilde{v}_{\bm{p}}(\eta) &=&  \left(\frac{\eta}{\eta_*}\right)^{-\nu} \hat{p}_i\Big[ v^i_{\bm{p}} e^{-ic_s p\eta} + v^{*i}_{-\bm{p}} e^{ic_s p\eta}  \Big], \label{eq:sw_v}\\
    \tilde{\lambda}_{\bm{p}}(\eta) &=& -\frac{i}{c_s} \left(\frac{\eta}{\eta_*}\right)^{-\nu} \hat{p}_i\Big[ v^i_{\bm{p}} e^{-ic_s p\eta} - v^{*i}_{-\bm{p}} e^{i c_s p\eta} \Big],
\end{eqnarray}
\end{subequations}
with wave amplitude $v^i_{\bm{p}} = -\hat{p}^ie^{i\frac{\pi}{2}\nu}(c_2-ic_1)/2c_s$. Notice the distinction between the plane wave amplitudes $v^i_{\bm{p}}, v^{*\,i}_{\bm{p}}$ and the Fourier transform of the fluid variables  $\tv^i_{\bm{p}}, \tilde{\lambda}_{\bm{p}}$. 
The plane wave amplitudes specify the initial conditions at time $\eta_*$ when the sound waves start propagating freely and encode all the statistical properties of the velocity and density fields. For compressional modes, we have
\begin{equation}\label{eq:P}
    \langle   v_{\bm{p}_1}^i v_{\bm{p}_2}^{*\, j}\rangle = \hat{p}^i_1\hat{p}^j_1 P_v(p_1) (2\pi)^3 \delta^3(\bm{k}_1 - \bm{k}_2).
\end{equation}
In the original papers~\cite{Hindmarsh:2016lnk, Hindmarsh:2019phv}, it was argued that, due to the incoherent sum of contributions from individual bubbles that collide at different times, the cross-correlator $\langle v_{\bm{k}_1}^i v_{-\bm{k}_2}^{j}\rangle$ are exponentially suppressed with respect to the correlator~\eqref{eq:P}.
Neglecting these terms, we simplify the UETC $C_{\tilde{v}\tilde{v}}$ of Fourier modes~\eqref{eqs:GG} as
\begin{equation}\label{eq:Cvv}
    C_{\tilde{v}\tilde{v}}(p, \eta_1, \eta_2) = 2 \left(\frac{\eta_1 \eta_2}{\eta_*^2}\right)^{-\nu} P_v(p)  \cos(c_s p\eta_-),
\end{equation}
with $\eta_- = \eta_1 - \eta_2$. Finally, we can write the shear stress UETC~\eqref{eq:U} as
\begin{equation}\label{eq:Upi}
    U_{\mathcal{S}}(k, \eta_1, \eta_2) = \frac{1}{\pi^2 k} \left(\frac{\eta_*^2}{\eta_1 \eta_2}\right)^{2\nu} \int_0^\infty dp \int_{\vert p -k\vert}^{p+k} dq (1-\mu_p^2)^2 \frac{p^3}{q}  P_{v}(p) P_{v}(q) \cos(c_s p\eta_-) \cos(c_s q\eta_-).
\end{equation}

\subsection{The gravitational waves power spectrum}
Given the UETC of the shear stress~\eqref{eq:Upi}, we can finally evaluate the dimensionless power spectrum of gravitational waves~\eqref{P_hp}. Long after the end of the acoustic phase $\eta\gg \eta_{\text{end}}$ and after the return to radiation time $\eta > \eta_{\text{r}}$, we can write
\begin{eqnarray}\label{eq:Pgw}
    \Pgw =  3(1+\omega)^2  \left(\frac{a_*}{a_{\text{r}}}\right)^{\frac{2\nu}{1+\nu}} (\mathcal{H}_* \eta_*)^2 \frac{k^3}{2\pi^2}\frac{1}{\pi^2 k}\int_0^\infty dp \int_{\vert p - k\vert}^{p+k}dq \, (1-\mu_p^2)^2 \frac{p^3}{q} \times\nonumber\\
    \times P_v(p) P_v(q) \Delta(k, p, q, \eta, \eta_{*}, \eta_{\text{end}}),
\end{eqnarray}
where we defined the kernel
\begin{equation}\label{eq:delta_i}
    \Delta(k, p, \Tilde{p}, \eta, \eta_*, \eta_{\text{end}}) = \iint_{\eta_*}^{\eta_{\text{end}}} \frac{d\eta_1 d\eta_2}{\eta_*^2} \left(\frac{\eta_*^2}{\eta_1\eta_2}\right)^{1+\nu} G_{k}^\prime(\eta, \eta_1)G_{k}^\prime(\eta, \eta_2) \cos(c_s p\eta_-)\cos(c_s q\eta_-).
\end{equation}
The scale dependence of the power spectrum can be made explicit by rescaling the inverse length variables as $z=kR_*$, $x = pR_*$, $\Tilde{x} = qR_*$ and the time variables $\uptau \equiv\eta/R_*$, $\uptau_* \equiv \eta_*/R_*$, $\uptau_{\text{end}} \equiv \eta_{\text{end}}/R_*$. 
We further introduce a dimensionless spectral density $\tilde{P}_v(p)$ through 
\begin{equation}
    P_v(p) \equiv v_{\text{rms}}^2 R_*^3 \tilde{P}_v\left(p R_*\right),
\end{equation}
where $v_{\text{rms}}$ denotes the root mean square (RMS) fluid velocity. The gravitational wave power spectrum can now be written as
\begin{equation}\label{eq:pgw_1}
    \boxed{\Pgw =  3\left(\Gamma v_{\text{rms}}^2\right)^2 (\mathcal{H}_* R_*) (\mathcal{H}_* \eta_*) \left(\frac{a_*}{\bar{a}_{\text{r}}}\right)^{\frac{2\nu}{1+\nu}}  \frac{(kR_*)^3}{2\pi^2} \tilde{P}_{\text{gw}} (kR_*)}
\end{equation}
with adiabatic index $\Gamma = \bar{w}/\bar{e}$. We further defined the dimensionless gravitational wave spectral density function
\begin{equation}\label{eq:pgw}
    \tilde{P}_{\text{gw}} (kR_*) = \frac{\uptau_*}{\pi^2 z^3}  \int_0^\infty dx \int_{\vert x - z\vert}^{x+z}d\tilde{x} \, \rho(z, x, \tilde{x}) \tilde{P}_v(x) \tilde{P}_v(\tilde{x}) \Delta \left(z, x, \tilde{x}, \tau, \tau_*, \tau_{\text{end}}\right),
\end{equation}
with
\begin{equation}\label{eq:rho_1}
    \rho(z, x, \tilde{x}) = \frac{\left[ \tilde{x}^2 - (x-z)^2\right]^2\left[ (x+z)^2 -  \tilde{x}^2\right]^2}{16x\tilde{x}z^2} 
\end{equation}
a geometric function of wavenumbers only. The geometric function $\rho(z, x, \tilde{x})$ incorporates the projected sound wave wavenumbers onto the TT-subspace and vanishes at $\tilde{x} = |x-z|$ and $\tilde{x} = x+z$. Equation~\eqref{eq:pgw_1} recovers the result of Ref.~\cite{Hindmarsh:2015qta} in the limit $\nu\rightarrow 0$. We notice that equation~\eqref{eq:pgw_1} does not exhibit an explicit linear dependence on the acoustic source duration $\detav \equiv \eta_\text{end} - \eta_*$, as it was found instead in the case of a flat static Universe~\cite{Hindmarsh:2015qta, Hindmarsh:2016lnk, Hindmarsh:2017gnf, Hindmarsh:2019phv}. The dependence on $\detav$ is encoded in the kernel~\eqref{eq:delta_i}. Given that the source of shear-stress decorrelates rapidly with time~\cite{Guo:2020grp}, we expect the peak amplitude of the gravitational wave power spectrum to grow linearly with $\detav$ as long as $\detav \ll \eta_*$~\cite{RoperPol:2023dzg}. We finally remark that equation~\eqref{eq:pgw_1} ignores the evolution of the shear-stress UETC due to formation of shocks and turbulence in the fluid~\cite{Dahl:2021wyk, Dahl:2024eup}.

As the phase transition releases most of the kinetic energy at the peak frequency $k_p \sim R_*^{-1}$, and given the assumption that $\mathcal{H}_* R_* \ll  1$, the loudest gravitational wave signals are emitted at sub-horizon scales $k\eta_* \gg 1$. 
With the same level of approximation that we used in Section~\ref{sec:correlators},  we expand the Green's function~\eqref{eq:Green} at leading order in short wavelength of gravitational waves and find (the details of this calculation can be found in Appendix~\ref{app:Green_function})
\begin{equation}
    G_k(\eta, \eta_1) = \frac{1}{k}\sin\left[k(\eta-\eta_1)\right]\Theta(\eta - \eta_1) + O\left(\frac{1}{k\eta}, \frac{1}{k\eta_1}\right).
\end{equation}
This is nothing but the expression of the gravitational wave Green's function in a Universe dominated by radiation $\nu =0$~\cite{Hindmarsh:2019phv, RoperPol:2023dzg}. Corrections from the speed of sound are only relevant at next to leading order in $k\eta_*$, and will be considered in a future work.
We stress the fact that this approximation might fail to capture the effects of a softened EOS on large scales $(k\rightarrow 0)$. In this regime however, the shape of the spectrum is constrained by the causality of the fluid flow. Long after the dissipation of sound waves, at conformal time $\eta\gg \eta_{1}, \eta_2$, we can average the Green's functions over a large number of oscillations and, from equation~\eqref{eq:delta_i}, obtain
\begin{equation}\label{eq:kernel_one}
    \Delta(z, x, \Tilde{x}, \uptau_*, \uptau_{\text{end}}) = \frac{1}{2} \iint_{\uptau_*}^{\uptau_{\text{end}}}  \frac{d\uptau_1 d\uptau_2}{\uptau_*^2}  \left(\frac{\uptau_*^2}{\uptau_1 \uptau_2}\right)^{1+\nu}  \cos(z\uptau_-) \cos(c_s x\uptau_-)\cos(c_s \tilde{x}\uptau_-).
\end{equation}
We can now perform the integration over $\uptau_1$ and $\uptau_2$ analytically
---the interested reader can find some supplemental material in Appendix~\ref{app:kernel}--- to get 
\begin{equation}\label{eqs:kernel_omega}
    \Delta (z, x, \Tilde{x}, \uptau_*, \uptau_{\text{end}}) = \frac{1}{8} \sum_{m, n = \pm 1} \left\vert\omega_{mn} \uptau_*\right\vert^{2\nu} \left[\left(\operatorname{ci}_{-\nu}(\omega_{m n} \uptau)\Big\vert^{\uptau_{\text{end}}}_{\uptau_*} \right)^2  + \left( \operatorname{si}_{-\nu}(\omega_{m n} \uptau)\Big\vert^{\uptau_{\text{end}}}_{\uptau_*} \right)^2 \right]
\end{equation}
with
\begin{equation}
    \omega_{m n} = z +c_s(mx + n\Tilde{x}), \qquad \nu = \frac{1-3\omega}{1+3\omega},
\end{equation}
and 
\begin{equation}\label{eqs:trigo}
     \operatorname{si}_\nu(x) =  \int_x^\infty \frac{\sin(t)}{t^{1-\nu}}dt ,\qquad\quad \operatorname{ci}_\nu(x) =  \int_x^\infty \frac{\cos(t)}{t^{1-\nu}}dt,
\end{equation}
the generalized sine and cosine integral functions as defined in~\cite[\href{https://dlmf.nist.gov/8.21.E4}{(8.21.4)}]{NIST:DLMF} and~\cite[\href{https://dlmf.nist.gov/8.21.E5}{(8.21.5)}]{NIST:DLMF} respectively. We chose to use a more compact notation, so that $\operatorname{ci}_\nu(x) \equiv \operatorname{ci}(\nu,x)$ and $\operatorname{si}_\nu(x) \equiv \operatorname{si}(\nu,x)$. Notice that in the limit $\nu\rightarrow 0$, the sign convention is such that $\operatorname{ci}(0,x) = - \operatorname{Ci}(x)$ and $\operatorname{si}(0,x) = - \operatorname{si}(x)$.
For the sake of convenience, we also introduced a new notation $f(\uptau)\vert^{\uptau_{\text{end}}}_{\uptau_*} \equiv f(\uptau_{\text{end}}) - f(\uptau_*)$ for any function $f$.

\paragraph{The case of radiation} 
In a Universe dominated by ultra-relativistic species,  $\omega = 1/3$ and the conformal parameter $\nu = 0$, so that 
the kernel function~\eqref{eqs:kernel_omega} recovers the well known expression~\cite{RoperPol:2023dzg,Sharma:2023mao}
\begin{equation}\label{eq:delta_0_rad}
    \Delta (z, x, \Tilde{x}, \uptau_*, \uptau_{\text{end}}) = \frac{1}{8} \sum_{m, n = \pm 1} \left[\left(\operatorname{ci}(\omega_{m n} \uptau)\Big\vert^{\uptau_{\text{end}}}_{\uptau_*} \right)^2  + \left( \operatorname{si}(\omega_{m n} \uptau)\Big\vert^{\uptau_{\text{end}}}_{\uptau_*} \right)^2 \right].
\end{equation}
In the limit $k\eta_*\rightarrow \infty$ the sine and cosine integral functions can be approximated for large arguments as in~\eqref{eq:trigo_large}. If we further assume that the source dissipates fast, i.e. $\eta_{\text{end}} /\eta_*\sim 1 $, the kernel recovers the expression 
\begin{equation}
    \Delta_{\text{flat}} (z, x, \Tilde{x}, \uptau_*, \uptau_{\text{end}}) \simeq \frac{1}{2} \sum_{m, n = \pm 1} \frac{\sin^2\left[\omega_{mn} (\uptau_\text{end} - \uptau_*)/2\right]}{(\omega_{mn}\uptau_*)^2}
\end{equation}
that describes the interference between gravitational and sound waves in a flat Minkowski background. In this limit the periodic functions oscillate very fast, so that the dominant contribution to the gravitational wave power spectrum is to be expected from the case $m=n=-1$~\cite{Sharma:2023mao}.

\section{Spectral shape of the gravitational wave power spectrum}\label{sec:shape}

The spectral density $P_v(p)$~\eqref{eq:P} specifies the distribution of the sound wave kinetic energy over all wavenumbers $p$ at initial time $\eta_*$, and
can be computed via numerical simulations~\cite{Hindmarsh:2015qta} or estimated with semi-analytic models as the SSM~\cite{Hindmarsh:2016lnk, Hindmarsh:2019phv}. 
However, for the purpose of this article, whose aim is to investigate the signature of speed of sound on the gravitational wave power spectrum,  we will just consider an analytic function that captures the causal structure of the correlator~\eqref{eq:P}. We will then consider
\begin{equation}\label{eq:Pv}
    P_v(p) = 3\pi \frac{v_{\text{rms}}^2}{k_p^3}\frac{(p/k_p)^2}{1+ (p/k_p)^6},
\end{equation}
which is a physically well motivated spectrum for a longitudinal causal flow~\cite{Durrer:2003ja} in presence of shocks~\cite{Dahl:2021wyk}. Parseval's theorem sets the normalization 
\begin{equation}
    v_{\text{rms}}^2 = \int\frac{d^3\bm{p}}{(2\pi)^3}C_{\tilde{v}\tilde{v}}(p, \eta_*, \eta_*),
\end{equation}
where the correlator $C_{\tilde{v}\tilde{v}}(p, \eta_*, \eta_*)$ is given by equation~\eqref{eq:Cvv}.

\subsection{Low-frequency regime: causal tail of the spectrum}\label{sec:low}
In the limit $k\rightarrow 0$, the sound wave wavenumbers align $\tilde{\bm{x}} = \bm{x} -\bm{z} \rightarrow \bm{x}$, and the kernel~\eqref{eq:kernel_one} can be integrated to
\begin{eqnarray}\label{eq:ker_low_z0}
    \Delta_{\text{low}} (x, \uptau_*, \uptau_{\text{end}}) &\xrightarrow[k\rightarrow 0]{}&  \frac{1}{4\nu^2}\left[1 - \left(\frac{\uptau_*}{\uptau_{\text{end}}}\right)^\nu\right]^2 + \nonumber\\
    && + \frac{(2c_sx\uptau_*)^{2\nu}}{4} \left[\left(\operatorname{ci}_{-\nu}(2c_s x\uptau) \Big\vert^{\uptau_{\text{end}}}_{\uptau_*} \right)^2  + \left( \operatorname{si}_{-\nu}(2c_s x \uptau)\Big\vert^{\uptau_{\text{end}}}_{\uptau_*} \right)^2 \right].\qquad\label{eq:delta_k0}
\end{eqnarray}
In the conformal limit $\nu \rightarrow 0$,  the first term  manifests a logarithmic behavior $1-(\eta_*/\eta_{\text{end}})^\nu \rightarrow - \nu \ln(\eta_*/\eta_{\text{end}})$, and the kernel becomes, as previously found in  Ref.~\cite{RoperPol:2023dzg},
\begin{equation}\label{delta_nu0}
    \Delta^{\nu = 0}_{\text{low}} (x, \uptau_*, \uptau_{\text{end}})  \xrightarrow[k\rightarrow 0]{}  \frac{1}{4}\ln^2\left(\frac{\uptau_{\text{end}}}{\uptau_*}\right) + \frac{1}{4}\left[\left( \operatorname{ci}(2c_s x\uptau)\Big\vert^{\uptau_{\text{end}}}_{\uptau_*} \right)^2  + \left( \operatorname{si}(2c_s x\uptau)\Big\vert^{\uptau_{\text{end}}}_{\uptau_*} \right)^2\right].
\end{equation}
The trigonometric integral functions provide an oscillatory decaying contribution that becomes subdominant when the source lasts for many Hubble times $\uptau_{\text{end}} \gg \uptau_*$. In the limit of infinite source duration $\uptau_{\text{end}}/\uptau_* \rightarrow \infty$, the kernel increases logarithmically with no bounds with the duration of the source in the case of radiation $(\nu=0)$, while it increases with asymptote $\Delta_{\text{low}}\rightarrow 1/4\nu^2$ whenever $\nu\neq 0$.

Since the kernel~\eqref{eq:ker_low_z0} does not depend on the sound wave scaled wavenumber $\tilde{x}$, we can separate the integration variables in equation~\eqref{eq:pgw} and to carry out one integration analytically. Using $d\tilde{x} \, \rho(z, x, \tilde{x})  = d\mu \,(1-\mu^2)^2 x^4 z^3 /\tilde{x}^2$, and performing the integration over the polar angle $\mu$, we find
\begin{equation}\label{eq:spec_low}
    \tilde{P}_{\text{gw}}^{\text{low}} (kR_*) =  \frac{16 \uptau_*}{15 \pi^2} \int_0^\infty dx x^2 \tilde{P}_v^2(x) \Delta_{\text{low}} (x, \uptau_*, \uptau_{\text{end}}).
\end{equation}
For long enough source duration, such that the contribution from the oscillatory functions in the kernel~\eqref{eq:ker_low_z0} become subdominant, we can simplify
\begin{equation}\label{eq:spec_low_long}
    \tilde{P}_{\text{gw}}^{\text{low}} (kR_*) =  \frac{1}{\nu^2}\left[1 - \left(1+\frac{\detav}{\eta_*}\right)^{-\nu}\right]^2 \frac{8}{15} \uptau_*\, \mathcal{I}_v,
\end{equation}
with $\detav \equiv \eta_{\text{end}}-\eta_*$ and
\begin{equation}\label{eq:I}
    \mathcal{I}_v \equiv \frac{1}{2\pi^2} \int_0^\infty dx x^2 \tilde{P}_v^2(x).
\end{equation}
The function $\mathcal{I}_v$ specifies the contribution from the source, and holds for a generic non relativistic, stationary and longitudinal flow. With the particular choice of spectral density~\eqref{eq:Pv}, we obtain $\mathcal{I}_v = 1/32\pi^2$.
At the level of the spectrum~\eqref{eq:pgw_1}, we confirm that, at length scales larger than the Hubble length $k\eta_* \ll 1$, causality enforces a universal profile $\Pgw(k) \propto k^3$ independently of the EOS \cite{Durrer:2003ja}.

\subsection{Intermediate-frequency regime: shallow growth}\label{subsec:intermediate}
In the intermediate frequency range, where the modes are generated with wavelengths smaller than the Hubble length, i.e. $k\eta_* \gg 1$, but much larger than the typical length scale of sound waves, i.e. $k\ll k_p$, the power spectrum of gravitational wave is dominated by the odd terms in the kernel~\eqref{eqs:kernel_omega}, that is the terms of the sum with $mn=-1$~\cite{Sharma:2023mao}.  At leading order in $1/k\eta_*$ (a detailed derivation can be found in Appendix~\ref{app:kernel})
\begin{eqnarray}
    \Delta_{\text{int}}(z,\mu,\uptau_*, \uptau_{\text{end}}) & \underset{1\ll k\eta_* \ll k_p\eta_*}{\simeq}& \frac{1}{8} \sum_{\pm}\left(k\eta_*(1\pm \mu c_s)\right)^{-2} \Bigg[ 1 + \left(1+\frac{\detav}{\eta_*}\right)^{-2(1+\nu)} - \nonumber\\
    &&\qquad\qquad  -2 \left(1+\frac{\detav}{\eta_*}\right)^{-(1+\nu)} \cos\left[k\detav (1\pm \mu c_s)\right]\Bigg],\label{eq:kernel_interm}
\end{eqnarray}
with $\mu = \hat{\bm{p}}\cdot\hat{\bm{k}}$ and $\detav = \eta_\text{end}-\eta_*$.
The behavior of the kernel in this frequency range depends on the duration of the source. In particular we distinguish three different sources: 
\begin{enumerate}[i)]
    \item Short-lasting source: a source that lasts around one period of sound wave oscillations $k_p\detav \sim 1$, but much less than a Hubble time $\detav\ll \eta_*$. Gravitational wave modes in the intermediate frequency range $1\ll k\eta_* \ll k_p\eta_*$ do not have time to oscillate once during the acoustic phase. We believe this is a very difficult scenario to realize in practice, as sound waves typically take several oscillations to dissipate the energy~\cite{Guo:2020grp}. Nonetheless, we include this case of study in our analysis for the sake of completeness. 
    \item Medium-lasting source: a source that lasts for many periods of sound wave oscillations $k_p\detav \gg 1$, but much less than a Hubble time $\detav\ll \eta_*$. Gravitational waves in the intermediate frequency range $1\ll k\eta_* \ll k_p\eta_*$ oscillate $k\detav\sim \mathcal{O}(1)$ times during the acoustic phase.
    \item Long-lasting source: a source that lasts much longer than a Hubble time $\detav\gg \eta_*$.
\end{enumerate}
For each case, in the regime $1\ll k\eta_* \ll k_p\eta_*$, we can approximate the kernel~\eqref{eq:kernel_interm} respectively as
\begin{subnumcases}{\Delta_{\text{int}}\sim }\text{i)} \;\;
  \frac{1}{4}\left(\frac{\detav}{\eta_*}\right)^2 & $\begin{array}{ccc} 
      \detav &\ll & \eta_*, \\
      k_p\detav& \sim & 1 .
    \end{array}$ \label{eq:kernel_short} \\    
  \noalign{\vskip5pt}
  \text{ii)} \; \, \Big[\frac{1}{2}- \frac{1+\nu}{2}\frac{\detav}{\eta_*}  \Big] \sum_{\pm}    \left(\frac{\sin\left(k\detav(1\pm \mu c_s)/2\right)}{k\eta_* (1\pm \mu c_s)}\right)^2 &  $ \begin{array}{ccc} 
      \detav &\ll & \eta_*, \\
      k_p\detav & \gg & 1. 
    \end{array}$ \label{eq:kernel_medium}\\  
  \noalign{\vskip5pt}
  \text{iii)}\; \frac{1}{4(k\eta_*)^2} \frac{1+c_s^2\mu^2}{(1-c_s^2\mu^2)^2} & $ \begin{array}{ccc} 
      \detav &\gg & \eta_*.
    \end{array}$  \label{eq:kernel_long}
\end{subnumcases}
The kernel in the intermediate frequency band~\eqref{eq:kernel_interm} does not depend on the sound wave momenta $x$ and $\tilde{x}$ separately, but only on the polar angle $\mu$. This allows us to simplify the dimensionless spectral density of gravitational waves~\eqref{eq:pgw}, that we write as 
\begin{equation}\label{eq:P_gw_interm}
    \tilde{P}^{\text{int}}_{\text{gw}} (kR_*) \simeq 2\uptau_*\mathcal{I}_v \int_{-1}^{1} d\mu \, (1-\mu^2)^2  \Delta_{\text{int}} (z, \mu, \uptau_*, \uptau_{\text{end}}),
\end{equation}
with $\mathcal{I}_v$ defined in equation~\eqref{eq:I}. Since the gravitational wave spectrum~\eqref{eq:pgw_1} scales as $\mathcal{P}_{\text{gw}}(k)\sim z^3 \tilde{P}_{\text{gw}}(z)$, equation~\eqref{eq:kernel_short} implies that, in the case of a short-lasting source, $\mathcal{P}_{\text{gw}}(k)$ grows quadratically with the source duration $\detav/\eta_*$ and cubically with the gravitational wave wavenumber $k$ in the intermediate frequency range $1\ll k\eta_* \ll k_p\eta_*$~\cite{RoperPol:2023dzg}. The causality tail $\mathcal{P}_{\text{gw}}(k)\propto k^3$ that characterizes the limit $k\rightarrow 0$ studied in Section~\ref{sec:low} extends, for short-lasting sources, to the peak $k\sim k_p$. 

The oscillatory behavior of the kernel becomes relevant for medium-lasting sources~\eqref{eq:kernel_medium} when $k_p \detav \gg 1$. If the source survives for several gravitational wave oscillations, the spectrum becomes insensitive to each individual oscillation and the kernel tends to equation~\eqref{eq:kernel_long}, which shows the emergence of a shallow spectrum  $\mathcal{P}_{\text{gw}}(k)\propto k$~\cite{Sharma:2023mao, RoperPol:2023dzg}.

The different approximations to $\Delta_{\text{int}}$ for short-, medium- and long-lasting sources, allow one to perform analytically the integration over $\mu$. In particular, for a long-lasting source, we find
\begin{equation}\label{eq:spec_int}
    \tilde{P}^{\text{int}}_{\text{gw}} (kR_*) \underset{\detav\gg \eta_*}{\simeq} \frac{4}{3c_s^4}  \left[ 3-2c_s^2 - \frac{3}{c_s}(1-c_s^2) \operatorname{arctanh}(c_s) \right]   \frac{\mathcal{I}_v}{\uptau_* z^2}.
\end{equation}

\subsection{High-frequency regime: spectral peak amplitude}\label{sec:high}
Modes with wavelength around the typical length scale of sound waves, $k \gtrsim k_p$, oscillate several times during the acoustic phase. Over many oscillations, the sound waves in the fluid and the emitted gravitational wave interfere constructively and destructively when the waves are out of phase. A strong resonance is found when the sound waves and the gravitational wave are in phase, that is when $z -c_s(x+\tilde{x}) = 0$~\cite{Hindmarsh:2019phv, Sharma:2023mao, RoperPol:2023dzg}. This justifies the approximation of the kernel~\eqref{eqs:kernel_omega} as a Dirac delta function centered at the resonance scale, that is (a rigorous derivation can be found in the Appendix~\ref{app:kernel})
\begin{equation}
    \Delta_{\text{high}}(z,x,\tilde{x},\uptau_*, \uptau_{\text{end}}) \simeq \frac{\pi}{4(1+2\nu)}\, \uptau_*^{-1} \left[1-\left(\frac{\eta_*}{\eta_{\text{end}}}\right)^{1+2\nu} \right] \delta\big( z -c_s(x + \tilde{x})\big).
\end{equation}
The integration over scaled momentum $\Tilde{x}$ in
the power spectrum~\eqref{eq:pgw} is now trivial, and can be performed analytically to get 
\begin{equation}\label{eq:pgw_final}
\begin{split}
    \Tilde{P}^{\text{high}}_{\text{gw}}(kR_*) &\simeq  \frac{1}{4\pi z c_s} \left(\frac{1-c_s^2}{c_s^2}\right)^2  \frac{1}{1+2\nu}\left[1-\left(1+\frac{\detav}{\eta_{*}}\right)^{-1-2\nu} \right] \times\qquad\qquad\qquad\qquad \\
    & \qquad\qquad\qquad\quad \times\int_{x_-}^{x_+}\frac{dx}{x} \frac{(x-x_+)^2(x-x_-)^2}{(x_+ + x_- - x)} \Tilde{P}_v(x) \Tilde{P}_v(x_+ + x_- -x),
\end{split}
\end{equation}
with $x_\pm = z(1\pm c_s)/(2c_s)$. Equations~\eqref{eq:pgw_final} and~\eqref{eq:pgw_1} extend the previous results on acoustically generated gravitational waves in the SSM~\cite{Hindmarsh:2019phv} to a wider class of EOS, and they correctly recover the expression of the gravitational wave power spectrum for stationary source in a Universe dominated by radiation by taking the conformal limit $\nu\rightarrow 0$~\cite{Hindmarsh:2015qta}. These expressions ignore however the shape evolution of the spectral density $P_v(p)$ due to the formation of shocks in the fluid as discussed in Refs.~\cite{Dahl:2021wyk, Dahl:2024eup}

For a source that lasts much less than a Hubble time we can approximate
\begin{equation}\label{eq:P_gw_short}
    \Tilde{P}^{\text{high}}_{\text{gw}}(kR_*) \underset{\detav\ll \eta_*}{\simeq}  \frac{1}{4\pi z c_s} \left(\frac{1-c_s^2}{c_s^2}\right)^2  \frac{\detav}{\eta_*}\int_{x_-}^{x_+}\frac{dx}{x} \frac{(x-x_+)^2(x-x_-)^2}{(x_+ + x_- - x)} \Tilde{P}_v(x) \Tilde{P}_v(x_+ + x_- -x),
\end{equation}
which shows that the amplitude of the gravitational wave power spectrum in the high frequency regime grows linearly with the source duration as long as $\detav \ll \eta_*$.

\paragraph{The profile around the peak}
The spectral density of velocity perturbations~\eqref{eq:Pv} is a steep function of comoving wavenumber $k$ with a sharp maximum around the peak scale $k_p$. For a rough estimate of the spectrum peak location and amplitude, we can approximate the function $x\Tilde{P}_v(x)$ as a Dirac delta function centered at its maximum\footnote{We prefer to extremize the function $x\Tilde{P}_v(x) (x_++x_--x)\Tilde{P}_v(x_++x_--x)$ and not simply $\Tilde{P}_v(x)\Tilde{P}_v(x_++x_--x)$ because with the former choice the geometric factor in the power spectrum~\eqref{eq:pgw_final} remains scale invariant and marginally contributes to the determination of the peak position.}~\cite{Sharma:2023mao}. With this approximation we find that maximum of the gravitational wave power spectrum is located at
\begin{equation}\label{eq:kstar}
    k_{\star} \simeq 2 c_s  k_p,
\end{equation}
and the amplitude of the spectrum $\mathcal{P}^{\star}_{\text{gw}} \equiv \mathcal{P}_{\text{gw}} (k_\star)$ can be roughly estimated as
\begin{equation}\label{eq:p_max}
    \mathcal{P}^{\star}_{\text{gw}} \simeq 3\left(\Gamma v_{\text{rms}}^2\right)^2 (\mathcal{H}_* R_*) (\mathcal{H}_* \eta_*) \left(\frac{a_*}{\bar{a}_{\text{r}}}\right)^{\frac{2\nu}{1+\nu}}  \left(\frac{1-c_s^2}{c_s^2}\right)^2 \frac{1}{1+2\nu}\left[1-\left(\frac{a_*}{a_{\text{end}}}\right)^{\frac{1+2\nu}{1+\nu}} \right]\frac{9}{(2\pi)^3}c_s^5.
\end{equation}
If the source lasts much less than a Hubble time, then 
\begin{equation}
    \Pgw^{\star} \underset{\detav\ll \eta_*}{\simeq}  3\left(\Gamma v_{\text{rms}}^2\right)^2 (\mathcal{H}_* R_*) (\mathcal{H}_* \detav) \left(\frac{a_*}{\bar{a}_{\text{r}}}\right)^{\frac{2\nu}{1+\nu}}  \left(\frac{1-c_s^2}{c_s^2}\right)^2 \frac{9}{(2\pi)^3}c_s^5.
\end{equation}
Since sound waves oscillate with a frequency that is directly proportional to the speed of sound $c_s$, the softening of the EOS shifts the spectrum peak amplitude towards smaller frequencies, as suggested by equation~\eqref{eq:kstar}. The linear relation with the speed of sound is indeed inherited by $k_\star$, which corresponds to the frequency at which the emitted gravitational wave resonates with the sound waves.

Given the shape of the velocity spectral density~\eqref{eq:Pv}, we can further infer the power-law indices of the spectrum~\eqref{eq:pgw_1} with a simple power counting. If the spectral density scales as $P_v\sim k^n$, with $n = 2$ for $k<k_\star$ and $n = -4$ for $k>k_\star$, then $\Tilde{P}_{\text{gw}}(kR_*)\sim k^{2n+2}$ and the total gravitational wave power spectrum
\begin{equation}\label{eq:power_law}
    \Pgw(k) \sim \left\lbrace\begin{array}{lcr}
        k^{9} & & k<k_\star, \\
        k^{-3} & & k>k_\star.
    \end{array} \right.
\end{equation}
The softening of the equation of state does not modify the power-law scaling  of the spectrum.

\section{Results and discussion: the gravitational wave power spectrum}\label{sec:results}

The shape and amplitude of the power spectrum~\eqref{eq:pgw_1} are controlled by five parameters: (i) the RMS fluid velocity $v_{\text{rms}}$; (ii) the characteristic length of sound waves $R_*$; (iii) the ratio $a_*/a_{\text{r}}$ of the scale factor evaluated at the beginning of the acoustic phase and at the return to the radiation dominated epoch;  (iv) the equation of state parameter $\omega$, which in our model determines uniquely the sound speed as $c_s^2 = \omega$; and (v) the  duration of the acoustic phase $\detav=\eta_\text{end} - \eta_*$. The latter can be measured in numbers $N_{\text{sh}}$ of shock formation times 
\begin{equation}
    \detav = N_{\text{sh}}\eta_{\text{sh}},
\end{equation}
where  
\begin{equation}
    \eta_{\text{sh}} \equiv \frac{\xi_*}{v_{\text{rms}}}
\end{equation}
is the typical timescale of formation and decay of shocks. Here, we choose to define $\eta_{\text{sh}}$ in terms of the integral scale $\xi_*$, derived from the fluid velocity by
\begin{equation}
    \xi_* \equiv \frac{1}{v_{\text{rms}}^2} \int\frac{d^3\bm{p}}{(2\pi)^3} p^{-1} P_v(p) = \frac{1}{4\pi\sqrt{3}} R_*.
\end{equation}
It determines the approximate length scale over which the sound waves are correlated and where most of the energy is located. 

The RMS fluid velocity controls the amplitude of the spectrum and is left as a free parameter of the model. We consider two benchmark values for the peak frequency of the sound waves $k_p\eta_* \in \{100, 200\}$; both values describe bubbles that were well inside the Hubble horizon at the beginning of the acoustic phase, so that higher order relativistic effects are subdominant. We consider three different values for the return to radiation time: a short, an intermediate, and a long soft phase with duration $a_*/a_{\text{r}} \in \{0.1,\, 0.5,\, 0.9\}$ respectively. The sound speed and the duration of the acoustic phase are instead varied discontinuously within their domains $c_s \in \, (0, \, 1/\sqrt{3}]$ and $N_{\text{sh}} \in \; (0, \infty )$.
Throughout this paper we always adopt the normalization $\eta_* = 1$.
The numerical integration of the dimensionless gravitational wave spectral density~\eqref{eq:pgw} is carried out by the \texttt{integrate.quad} routine of \texttt{SciPy}. In some cases, a special care is needed in the numerical evaluation of the kernel function to treat numerical divergences, whose details are described in Appendix~\ref{sec:append_numerics}.

\subsection{Time duration of the source}
\begin{figure}
    \centering
    \includegraphics[width=1.0\textwidth]{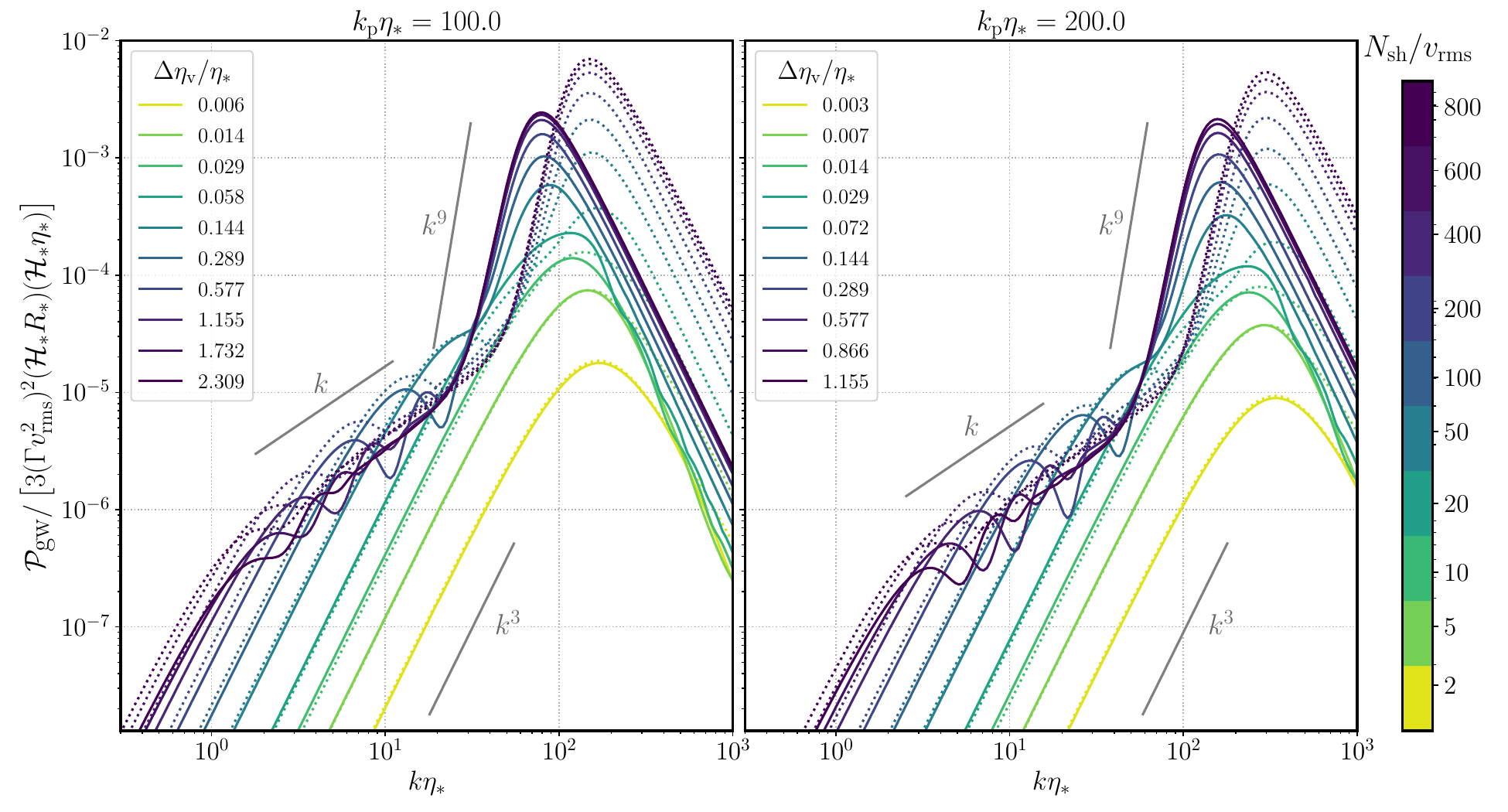}
    \caption{
    Gravitational waves power spectrum  $\Pgw$ for different source duration in the two benchmark cases $k_p\eta_* = 100$ (left) and $k_p\eta_* = 200$ (right). Solid lines correspond to $c_s = 0.31623$, dotted lines to $c_s = 1/\sqrt{3}$ (radiation). The radiation time has been fixed to  $a_*/a_r = 0.9$ in all the displayed cases. gray annotations indicate the power-law indices predicted by the SSM. Displayed are the cases of short-lasting sources with $N_{\text{sh}}/v_{\text{rms}} = [2,5,10,20]$, medium-lasting sources with $N_{\text{sh}}/v_{\text{rms}} = [50,100,200]$, and long-lasting sources with $N_{\text{sh}}/v_{\text{rms}} = [400, 600, 800]$.}
    \label{fig:convergence}
\end{figure}
In Figure~\ref{fig:convergence} we analyze the behavior of the power spectrum with the duration of the acoustic phase for the benchmark cases $k_p\eta_* \in \{100, 200\}$. As the source continuously injects energy into gravitational waves, the amplitude of the spectrum increases with the duration of the source. Short-lasting sources, that is the case when $\detav\ll \eta_*$ and when the gravitational wave modes with $k \ll k_p$ do not have time to complete one oscillation during the acoustic phase, are characterized by a causal tail $\Pgw\propto k^3$ that extend to the peak amplitude at $k_\star = 2c_s k_p$, as expected from~\eqref{eq:kernel_short}. 
Ahead of the peak, at frequencies $k>k_\star$, they show small oscillations; these frequencies indeed correspond to modes that oscillate $k\detav \gtrsim \mathcal{O}(1)$ times during the acoustic phase. 
As we increase the source duration to $k_p\detav \gg 1$ (medium-lasting sources), the spectrum starts  exhibiting a pronounced peak with a steep $k^9$ profile on the low-frequency side. At the intermediate frequency range $1\ll k\eta_* \ll k_p\eta_*$, the spectrum becomes shallower and develops oscillations, in agreement with the analytic result~\eqref{eq:kernel_medium}.  Increasing further the source duration to the limit $\detav \gg \eta_*$ (long-lasting sources), the spectrum in the intermediate frequency band converges to a shallow slope $\mathcal{P}_{\text{gw}} \propto k$, as in equation~\eqref{eq:kernel_long}.

Figure~\ref{fig:convergence} shows further that the spectrum always convergences to a limiting profile as $\detav/\eta_* \rightarrow\infty$ in the intermediate and high frequency regime, where $k\eta_* \gg 1$. Indeed, according to our analytic estimates~\eqref{eq:P_gw_interm} and~\eqref{eq:pgw_final}
\begin{subnumcases} {\mathcal{P}_{\text{gw}}(k) \underset{\detav \gg \eta_*}{\propto} \frac{(kR_*)^3}{2\pi^2}} 
        2\uptau_* \mathcal{I}_v \frac{1}{(k\eta_*)^2}, & $1\ll k\eta_* \ll k_p \eta_* $, \\
         \begin{aligned}
             &\frac{1}{z^3}\left[1-\left(\frac{\detav}{\eta_*}\right)^{-1-2\nu} \right]\int_{x_-}^{x_+} dx\,\rho(z, x) \times  \\
             &\qquad\qquad\qquad\quad\times  \Tilde{P}_v(x) \Tilde{P}_v(x_+ + x_- -x)
         \end{aligned},  & $k\gtrsim \mathcal{O}(k_p)$,   
\end{subnumcases}
with $\rho(z,x)$ defined in~\eqref{eq:rho(z,x)}.
This shows that the dimensionless spectral density function~\eqref{eq:pgw}, and thereby the power spectrum $\mathcal{P}_{\text{gw}}$, converges in the limit $\detav/\eta_* \rightarrow \infty$ regardless of the speed of sound. 

In the low frequency regime $k\eta_* \lesssim \mathcal{O}(1)$ instead, the convergence of the causality tails depends on the EOS. In particular we find that the convergence is achieved whenever $\nu \neq 0$, but never in an exact radiation era $(\nu = 0)$. 
This behavior is again well motivated by our analytic estimate~\eqref{eq:delta_k0},
\begin{eqnarray}
    \Delta (z, x, \Tilde{x}, \uptau_*, \uptau_{\text{end}}) &\xrightarrow[k\rightarrow 0]{}&  \frac{1}{4\nu^2}\left[1 - \left(\frac{a_*}{a_{\text{end}}}\right)^{\frac{\nu}{1+\nu}}\right]^2 ,
\end{eqnarray}
for which the dimensionless spectral density~\eqref{eq:P_gw_interm} can be integrated to 
\begin{equation}
    \tilde{P}_{\text{gw}} (kR_*) \xrightarrow[k\rightarrow 0]{}\frac{\uptau_*}{15\pi^2}   \frac{1}{\nu^2}\left[1 - \left(\frac{a_*}{a_{\text{end}}}\right)^{\frac{\nu}{1+\nu}}\right]^2.
\end{equation}
When $\nu \neq 0$, the spectral density increases as $\eta_\text{end}/\eta_* \rightarrow \infty$ with asymptote $\tilde{P}_{\text{gw}}^{\text{max}} = \uptau_*^2/(15\pi^2 \nu^2)$. On the contrary, when $\nu = 0$ the spectral density increases logarithmically with no bounds  as $\tilde{P}_{\text{gw}} \sim \uptau_* (1+\nu)^{-2}\ln^2(a_\text{end}/a_*)/ (15\pi^2) $.

\subsection{The effect of acoustic friction on the gravitational wave power spectrum}
\begin{figure}
    \centering
    \includegraphics[width=1.0\textwidth]{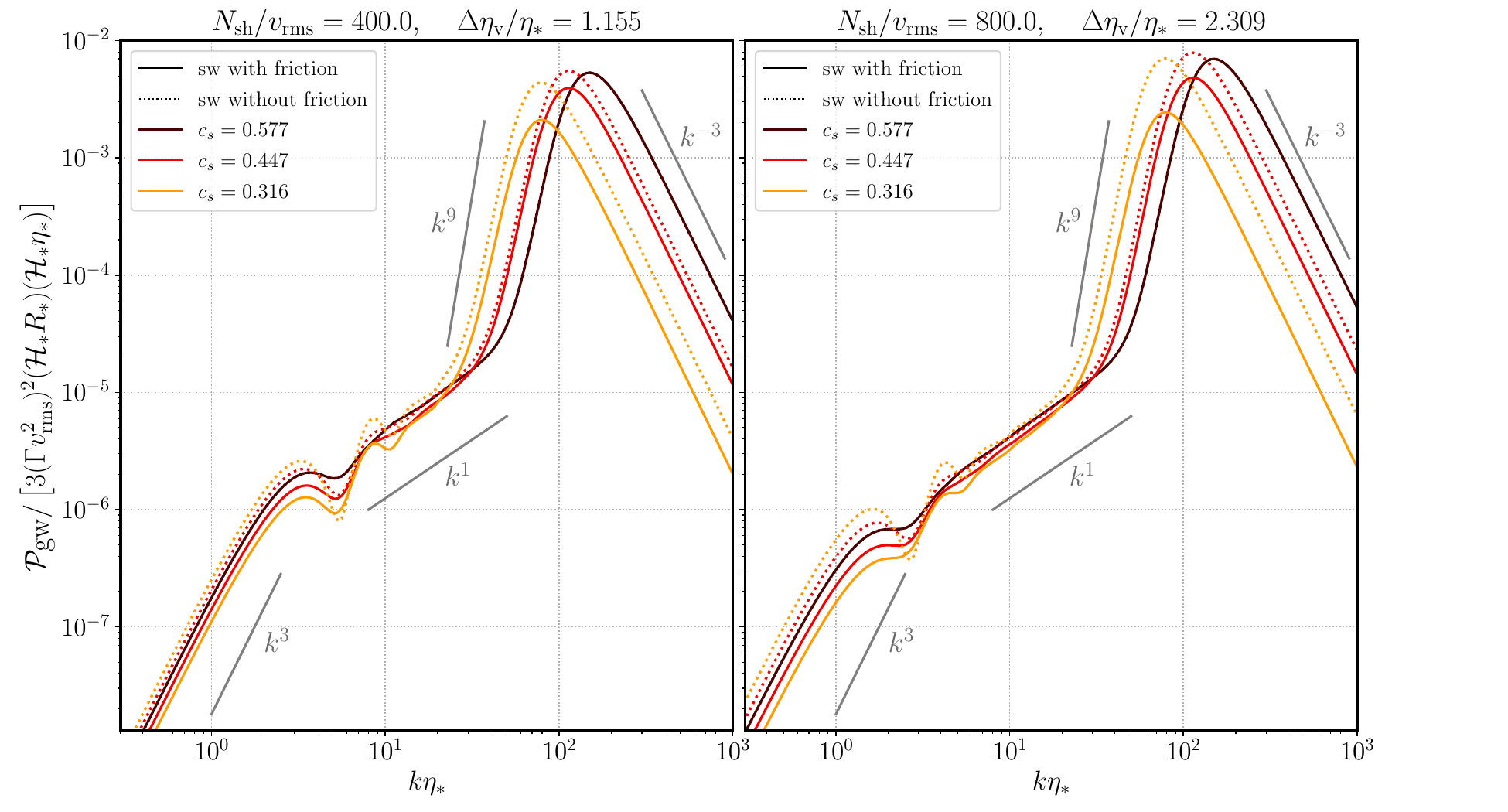}
    \caption{Implication of the sound wave friction. Comparison between the gravitational wave power spectrum computed with (solid lines) or without (dotted lines) the Hubble suppression in the propagation of sound wave. The characteristic length scale of sound waves is set to $k_p\eta_* = 100$, and the radiation time to $a_*/a_r = 0.9$. Dashed lines represent the analytic approximation~\eqref{eq:pgw_final} to the power spectrum. The gray solid lines highlight the power-law scaling predicted by the SSM.}
    \label{fig:decay}
\end{figure}

The effect of acoustic friction in the propagation of sound waves~\eqref{eq:sound_waves} is crucial to understand the role of the speed of sound on the gravitational wave power spectrum. 
In Figure~\ref{fig:decay} we compare the total gravitational wave power spectrum computed with or without the inclusion of the acoustic friction factor $(\eta_*/\eta)^\nu$ in equation~\eqref{eq:sound_waves}.
In the no-friction case, the shear-stress UETC~\eqref{eq:Upi} is not suppressed by the factor $(\eta_*^2/\eta_1\eta_2)^{2\nu}$, and the kernel~\eqref{eq:kernel_one} becomes
\begin{equation}\label{eq:kernel_nf}
    \Delta^{\text{nf}}(z,x,\tilde{x},\uptau_*, \uptau_\text{end}) = \frac{1}{2} \iint_{\uptau_*}^{\uptau_{\text{end}}}  \frac{d\uptau_1 d\uptau_2}{\uptau_*^2}  \left(\frac{\uptau_*^2}{\uptau_1\uptau_2}\right)^{1-\nu}  \cos(z\uptau_-) \cos(c_s x\uptau_-)\cos(c_s \tilde{x}\uptau_-).
\end{equation}
Taking the limit $\nu\rightarrow 0$ of equation~\eqref{eq:kernel_one} is not sufficient to describe the no-friction case, since an explicit factor $(\uptau_*^2/\uptau_1\uptau_2)^{-\nu}$ in the kernel comes directly from the background energy density of the cosmic fluid, as shown in equation~\eqref{P_hp}.
Figure~\ref{fig:decay} shows that neglecting friction can lead in general to a significant overestimation of the gravitational wave power spectrum. Comparing the no-friction kernel~\eqref{eq:kernel_nf} with equations~\eqref{eq:delta_hf} and~\eqref{eq:D}, we obtain an estimate of the total suppression carried by the sound wave friction on the peak amplitude of the power spectrum, that is
\begin{equation}\label{eq:math_f}
    \mathcal{F} \equiv \frac{\Delta}{\Delta^{\text{nf}}} \simeq \frac{1-2\nu}{1+2\nu} \frac{\left[1 - \left(1+\detav/\eta_*\right)^{-1-2\nu}\right]}{\left[1 - \left(1+\detav/\eta_*\right)^{-1+2\nu}\right]}.
\end{equation}
We notice that $\mathcal{F}$ is always positive and tends to $\mathcal{F} \rightarrow \frac{1}{2} \left[1 - \left(1+\detav/\eta_*\right)^{2}\right] \ln^{-1}\left(1+\detav/\eta_*\right)$ as \mbox{$\nu \rightarrow 1/2$}.

The power-law indices of the gravitational wave spectrum are not affected by the EOS, and agree with the power-laws~\eqref{eq:power_law} proposed by the SSM and its later developments~\cite{Hindmarsh:2019phv, RoperPol:2023dzg} irrespectively of $\omega$. 
Finally we appreciate the importance of the analytic result~\eqref{eq:pgw_final}, which provides a very good approximation of the power spectrum at its peak and in all the frequency region $k\gtrsim c_s k_p$.

\subsection{Suppression of gravitational waves}
\begin{figure}
    \centering
    \includegraphics[width=1.0\textwidth]{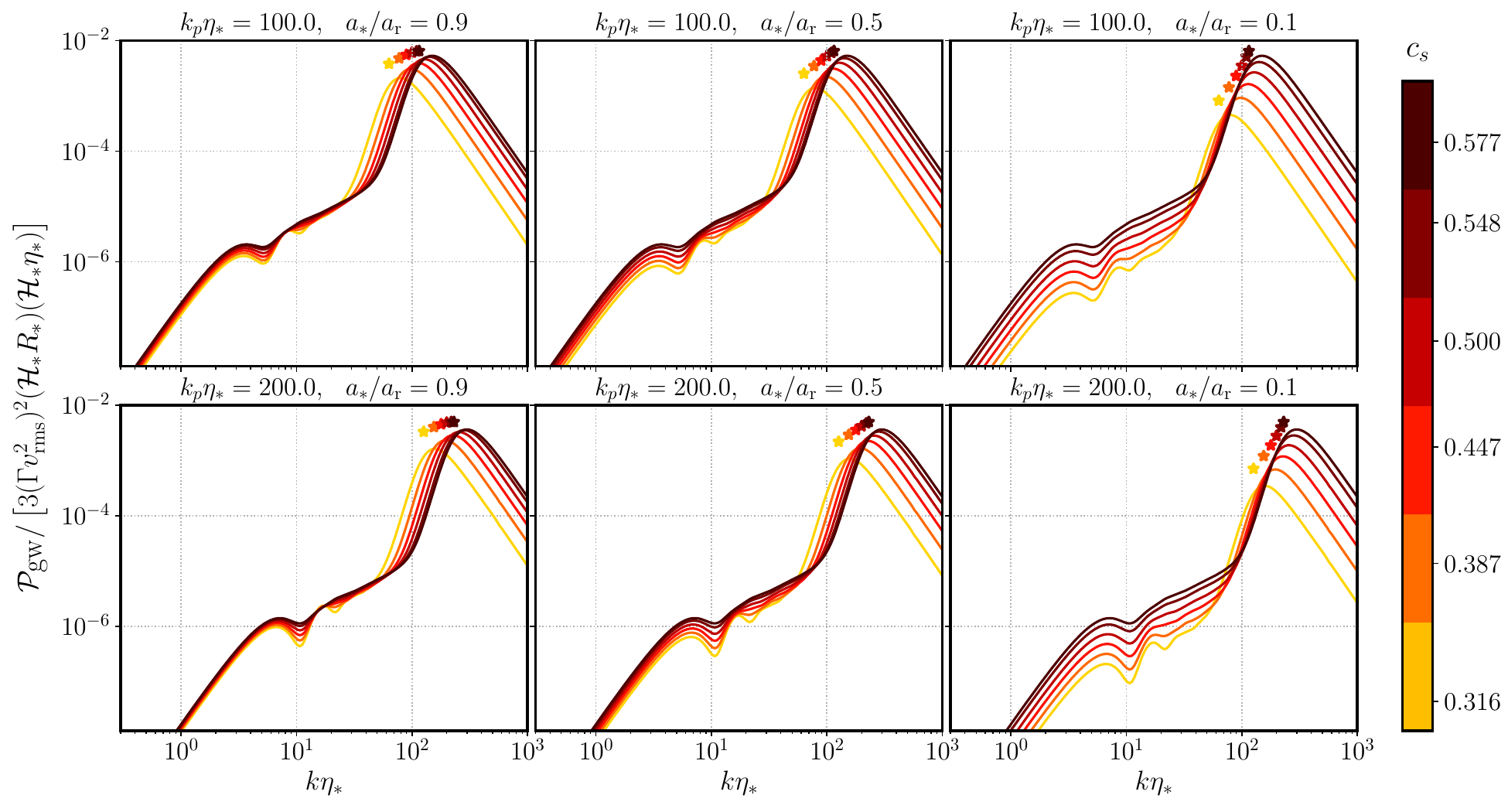}
    \caption{Gravitational waves power spectrum  for different values of the ratio $(a_*/a_\text{r})$ and different values of the sound speed $c_s$. The duration of the acoustic phase is fixed at $\Nsh/\vrms = 400$. The star-shaped markers indicate the frequency location $k_\star =2c_sk_p$ and amplitude of spectrum peak as estimated by our analytical approximation~\eqref{eq:p_max}.}
    \label{fig:scale_factor}
\end{figure}

As the equation of state gets softer, the contribution of pressure to the gravitational attraction becomes smaller and the Universe expands faster. This slows down the dilution of the background energy density of the Universe during the soft phase, and consequently suppresses the background energy density $\Bar{e}_*$ at the beginning of the acoustic phase with respect to the critical energy density $\bar{e}$ after the transition. A less energetic background is less efficient in fueling shear stress and sourcing gravitational waves, which then perceive a scale-independent suppression as in equation~\eqref{eq:a_ar}. 
We analyze this effect in Figure~\ref{fig:scale_factor},
where the total gravitational wave power spectrum is plotted for different values of the ratio $(a_*/a_\text{r})$. The softer the EOS and the longer the duration of the soft phase, the stronger the suppression of gravitational waves.

\subsection{Expression for the universal shape of the power spectrum}

\begin{figure}[t]
    \centering
    \includegraphics[width=1.0\textwidth]{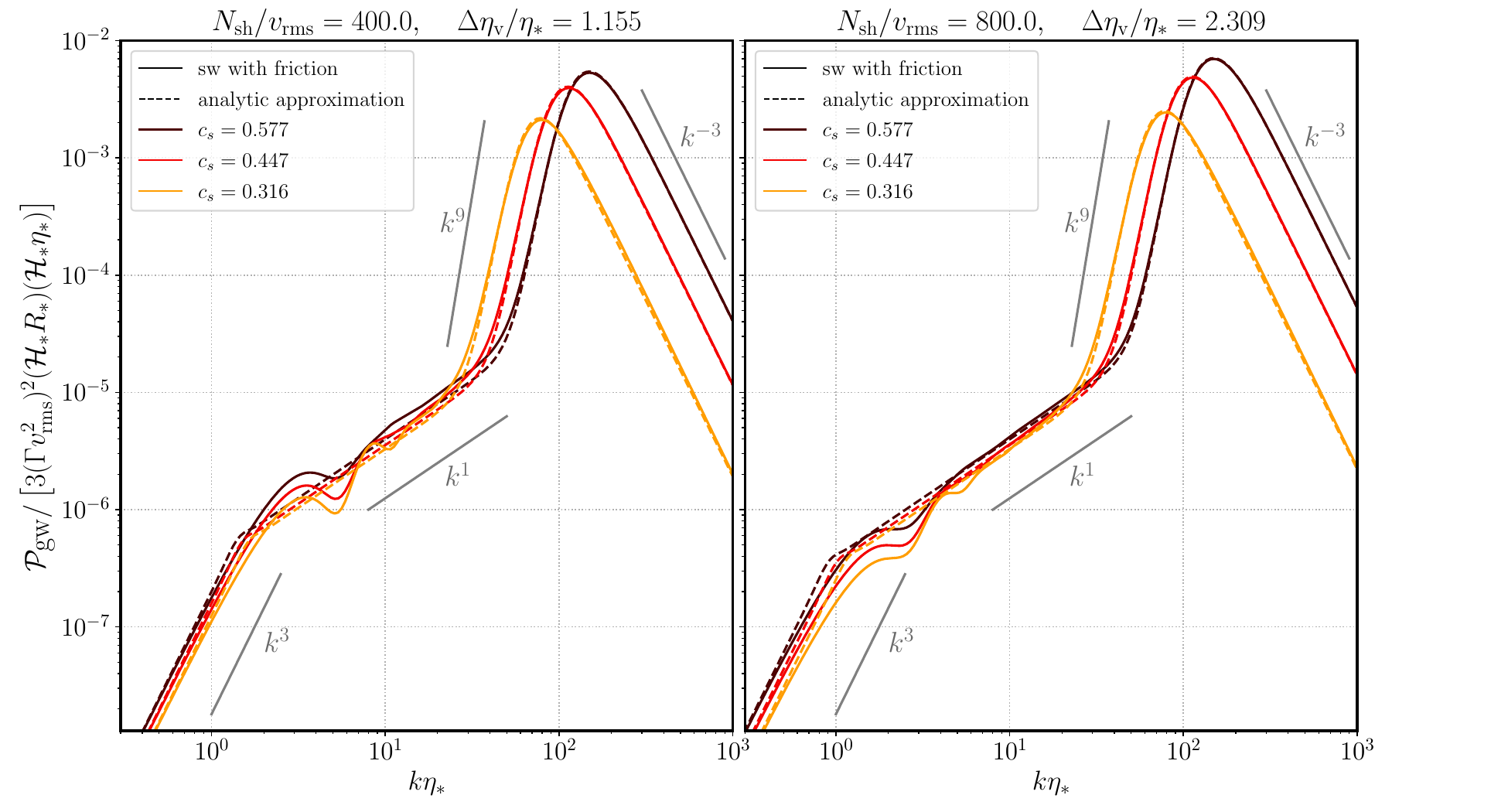}
    \caption{Comparison between the numerical results (solid line) on the gravitational wave power spectrum and its analytic approximation (dashed lines) as proposed in equation~\eqref{eq:global_approx}. The characteristic length scale of sound waves is set to $k_p\eta_* = 100$, and the radiation time to $a_*/a_r = 0.9$. The gray solid lines highlight the power-law scaling predicted by the SSM.}
    \label{fig:global}
\end{figure}

From a phenomenological point of view, it is useful to have a fast and simple way to estimate the gravitational wave power spectrum~\eqref{eq:pgw} at all scales. While the analytic integration of the power spectrum remains in general an unsolved challenge, we provide here a rough-and-ready recipe to approximate the spectrum across all scales. We propose
\begin{equation}\label{eq:global_approx}
\begin{split}
    \Pgw(k) = &  3\left(\Gamma v_{\text{rms}}^2\right)^2 (\mathcal{H}_* R_*) (\mathcal{H}_* \eta_*) \left(\frac{a_*}{\bar{a}_{\text{r}}}\right)^{\frac{2\nu}{1+\nu}}  \frac{(kR_*)^3}{2\pi^2} \times  \bigg\{ \tilde{P}_{\text{gw}}^{\text{low}} (kR_*)\frac{1}{2}\operatorname{erfc}\big(2\pi \eta_*(k-k_\times)\big)\quad\\
    & + \frac{1}{2}\big[1+\operatorname{erf}\big(2\pi \eta_*(k-k_\times)\big)\big] \,\tilde{P}_{\text{gw}}^{\text{int}} (kR_*) \frac{1}{2}\operatorname{erfc}\big(2\pi \eta_*(k-k_\star)\big) + \tilde{P}_{\text{gw}}^{\text{high}} (kR_*)\bigg\}\quad
\end{split}
\end{equation}
with $\tilde{P}_{\text{gw}}^{\text{low}}$ given by equation~\eqref{eq:spec_low}, $\tilde{P}_{\text{gw}}^{\text{int}}$ by equation~\eqref{eq:P_gw_interm} (or, in the case of long-lasting source, by equations~\eqref{eq:spec_low_long} and~\eqref{eq:spec_int} respectively), and $\tilde{P}_{\text{gw}}^{\text{high}}$ by equation~\eqref{eq:pgw_final}. The frequency $k_\star = 2c_s k_p$ corresponds approximately to the location of the peak amplitude, while the frequency $k_\times$ represents the scale where the spectrum smoothly switches between the low and intermediate frequency range.
We evaluated $k_\times$ as the root of $\tilde{P}_{\text{gw}}^{\text{low}}(k_\times R_*) = \tilde{P}_{\text{gw}}^{\text{int}}(k_\times R_*)$ that, for long lasting source, has solution
\begin{equation}
    k_\times = \frac{\sqrt{5}}{\sqrt{2} c_s^2} \frac{\nu}{1+\nu}\frac{\sqrt{ 3-2c_s^2 - \frac{3}{c_s}(1-c_s^2) \operatorname{arctanh}(c_s) }}{1 -\left(1+\frac{\detav}{\eta_*} \right)^{-\nu}}\; \mathcal{H}_*.
\end{equation}
The complementary error function $\operatorname{erfc}(x)$ turns off the contribution of $\tilde{P}_{\text{gw}}^{\text{low}}$ when $k>k_\times$ and the contribution of $\tilde{P}_{\text{gw}}^{\text{int}}$ when $k>k_\star$; the error function $\operatorname{erf}(x)$ instead turns on $\tilde{P}_{\text{gw}}^{\text{int}}$ in the intermediate frequency range when $k>k_\times$. The high-frequency term does not need a switch, because the dimensionless spectral density grows fast as $\tilde{P}_{\text{gw}}^{\text{high}} \propto k^6$ on the left side of the peak, so that this contribution becomes rapidly subdominant on small frequencies. 
An example of the whole spectrum approximation for the case of a long-lasting source is shown in Figure~\eqref{fig:global}.

\section{Conclusion}
The expansion rate of the Universe and the softening of the EOS are two important mechanisms that can significantly modify the shape of the power spectrum of acoustically generated gravitational waves. We have shown that the softening of the EOS decreases the typical frequency of sound waves, and shifts the location of spectrum peak amplitude $k_\star\simeq 2c_s k_p$ to smaller frequencies. Additionally, the softening of the EOS affects the energy density of gravitational waves in two ways: (i) it introduces a time-damping effect in the propagation of sound waves which suppresses the peak amplitude of the gravitational wave power spectrum by a factor $\mathcal{F}(\nu, \eta_*, \eta_\text{end})$ as in equation~\eqref{eq:math_f}, where $\nu$ parametrizes the deviation from a radiation equation of state; (ii) it decreases the background energy density of the cosmic fluid that sources gravitational waves compared to the critical density today, bringing an additional suppression by a factor $(a_*/a)^4 \, \bar{e}_*/\bar{e}$. The combination of these two effects on the peak amplitude of the gravitational wave power spectrum can be estimated as a suppression, with respect to the case of radiation $(\nu = 0)$~\cite{Hindmarsh:2013xza, Hindmarsh:2015qta, Hindmarsh:2016lnk, Hindmarsh:2017gnf, Hindmarsh:2019phv, RoperPol:2023dzg, Sharma:2023mao}, by a factor
\begin{equation}
    \frac{\mathcal{P}^{\star}_{\text{gw}}}{\mathcal{P}^{\star}_{\text{gw}, \,\nu=0}} \simeq \frac{1}{1+2\nu}  \frac{\left[1 - \left(a_*/a_{\text{end}}\right)^{\frac{1+2\nu}{1+\nu}}\right]}{\left[1 - \left(a_*/a_{\text{end}}\right)^{\frac{1}{1+\nu}}\right]}\left(\frac{a_*}{a_{\text{r}}}\right)^{2\nu/(1+\nu)},
\end{equation}
with $\mathcal{P}^{\star}_{\text{gw}} \equiv \mathcal{P}_{\text{gw}}(k_\star)$ the spectral peak amplitude, and $a_{\text{r}}$ the scale factor at the radiation time $\eta_{\text{r}}$ when the Universe transitions to a radiation epoch. 
The shape of the spectrum around the frequency peak is well captured by our analytic approximation~\eqref{eq:pgw_final}; this can be used as a template for many models of phase transitions when the cosmic fluid is described by a generic barotropic EOS. We remark that these results were found under the assumption that $R_*\mathcal{H}_*\ll 1$ and that the fluid shear-stress, the source of gravitational waves, is stationary and dominated by non-relativistic compressional modes.

The duration of the source affects the gravitational wave power spectrum differently across different frequency regimes. The power spectrum of low-frequency modes with $k\eta_* \ll 1$ grows quadratically with the source duration $\detav$ as long as $\detav \ll \eta_*$. For long-lasting sources $\detav \gg \eta_*$ we find that the low-frequency spectrum converges for every value of the EOS parameter but $\nu =0$, in which case the profile grows logarithmically with the source duration ~\cite{RoperPol:2023dzg}. In this frequency regime the gravitational wave spectrum always follow the causal profile $\mathcal{P}_{\text{gw}}\propto k^3$. In the intermediate frequency range $1\ll k\eta_* \ll k_p\eta_*$ the spectrum still grows quadratically with $\detav$ when $\detav \ll \eta_*$. Increasing the source duration, the intermediate-frequency spectrum develops oscillations when the source last for a number $k\detav\sim \mathcal{O}(1)$ of gravitational wave periods and converges to a shallow profile $\mathcal{P}_{\text{gw}}\propto k$ when $k\detav \gg 1$~\cite{Sharma:2023mao}. The spectrum of high-frequency modes around the peak $k\gtrsim k_p$ instead initially grows linearly with $\detav$ when $\detav\ll \eta_*$, and converges to a limiting profile when $\detav\gg \eta_*$~\cite{Hindmarsh:2019phv}.

This work expands the possible scenarios of cosmological phase transition that will be possible to test with future interferometers like LISA adding the contribution from a softer EOS. The model considers the source of shear stress to be approximately constant for a finite time. In a realistic scenario, the source evolves in time until sound waves decorrelate, leading to a time-dependent spectral density of fluid perturbation and time-dependent peak-frequency $k_p$~\cite{Dahl:2021wyk, Dahl:2024eup}. The computation of a time-evolving spectrum cannot be carried out analytically, and a more numerically demanding simulation is needed to perform the numerical integration of the spectrum.

A realistic model of the source, which in this work was approximated with an analytic function as in equation~\eqref{eq:Pv}, must also take into account the particular fluid dynamics around expanding bubbles, as done for example in the Sound Shell Model~\cite{Hindmarsh:2016lnk, Hindmarsh:2019phv}. The effects of the Universe expansion and self-gravitation of the fluid on the bubble dynamics have been investigated in Refs.~\cite{Cai:2018teh, Giombi:2023jqq, Jinno:2024nwb}, but still need to be studied at the level of the gravitational wave power spectrum.

At leading order in $R_*\mathcal{H}_* \ll 1$, general relativistic corrections on the gravitational wave power spectrum only introduce the acoustic friction in the propagation of sound waves that has been discussed in this work. 
Contributions beyond the leading order become important when the typical bubble size becomes comparable with the Hubble radius. These contributions affect the propagation of sound waves, as suggested already by equation~\eqref{eq:eom}, and gravitational waves as in equation~\eqref{eq:Green}. Moreover, scalar perturbations provide an additional contribution to the shear stress that sources gravitational wave whose impact has not been considered yet in the estimation of the total power spectrum. We reserve the analysis of these contributions to future studies.

\section*{Acknowledgements}
The work of LG has been funded by the Research Council of Finland grant numbers 333609 and 349865. MH was supported by the Research Council of Finland grant numbers 333609, and JD by the Research Council of Finland grants numbers 354572 and 353131.

\appendix
\section{Green's function solution}\label{app:Green_function}
The Green's function for a linear differential operator $\mathcal{L}(k,\eta)$ is obtained by combining 
two independent solutions $g_{k}(\eta)$ and $f_{k}(\eta)$ of the homogeneous equation $\mathcal{L}(k,\eta)g_{k}(\eta) = \mathcal{L}(k,\eta)f_{k}(\eta) =0$ as
\begin{equation}
    G_k(\eta, \eta_1) = \frac{1}{\mathcal{N}_k} \left[g_{k}(\eta)f_{k}(\eta_1) - g_{k}(\eta_1)f_{k}(\eta)\right] \Theta(\eta - \eta_1),
\end{equation}
with $\mathcal{N}_k = g_{k}^\prime(\eta_1) f_{k}(\eta_1) - g_{k}(\eta_1) f_{k}^\prime(\eta_1)$ 
a normalization that only depends on the time variable $\eta_1$, and $\Theta(\eta - \eta_1)$ the Heaviside step function. One can show that two homogeneous solutions of 
the wave equation~\eqref{eq:GW_rescaled} are $g_{k}(\eta) = k\eta\,j_{\nu}(k\eta)$ and $f_{k}(\eta) = k\eta\, y_{\nu}(k\eta)$, 
with $y_\nu$ and $j_\nu$ the spherical Bessel functions of first and second kind respectively of order $\nu = (1-3\omega)/(1+3\omega)$. Then
\begin{equation}\label{eq:green}
    G_k(\eta, \eta_1) = \frac{k^2}{\mathcal{N}_k}  \eta\eta_1 \left[j_{\nu}(k\eta)y_{\nu}(k\eta_1) - j_{\nu}(k\eta_1)y_{\nu} (k\eta) \right]\Theta(\eta - \eta_1), \qquad \nu =  \frac{1-3\omega}{1+3\omega}
\end{equation}
In the limit of large argument $k\eta \gg 1$, the spherical Bessel functions approximate to~\cite{ARFKEN2013643}
\begin{subequations}\label{eq:Bessel_limit}
\begin{eqnarray}
    j_{\nu}(k\eta) &\simeq& \frac{1}{k\eta} \cos\left[k\eta - \frac{\pi}{2} (\nu+1)\right] + \mathcal{O}\left( \frac{1}{(k\eta)^2}\right), \\
    y_{\nu}(k\eta) &\simeq& \frac{1}{k\eta} \sin\left[k\eta - \frac{\pi}{2} (\nu+1)\right] + \mathcal{O}\left( \frac{1}{(k\eta)^2}\right),
\end{eqnarray}
\end{subequations}
and the Green's function \eqref{eq:green} tends to the solution in an exact radiation era Universe
\begin{equation}\label{eq:Green2}
    G_k(\eta, \eta_1) = \frac{1}{k}\sin\left[k\left(\eta- \eta_1 \right)\right]\Theta(\eta - \eta_1) + \mathcal{O}\left(\frac{1}{(k\eta)^2}\right).
\end{equation}
In the expression~\eqref{eq:ll} for the two-point correlation function of gravitational waves, the Green's functions are integrated over the time duration of the source, and evaluated at much later time during the radiation dominated era. $\eta\gg \eta_{\text{end}}$.
Taking the first derivative with respect to conformal time of~\eqref{eq:Green2} and averaging the periodic functions over a large number of oscillations $(k\eta\gg 1)$, we get
\begin{eqnarray}\label{eq:gpgp_result}
    G_k^\prime(\eta, \eta_1)G_k^\prime(\eta, \eta_2) &=& \cos\left[k(\eta-\eta_1)\right]\cos\left[k(\eta-\eta_2)\right] \xrightarrow[\eta\gg\eta_{1}, \eta_2]{} \frac{1}{2}\cos\left[k\left(\eta_1 -\eta_2\right)\right].
\end{eqnarray}

\section{Supplemental material for the calculation of the gravitational wave power spectrum}\label{app:kernel}
In this section of the Appendix we elucidate the analytic calculation leading to the kernel function~\eqref{eqs:kernel_omega} as well as its approximations used in Section~\eqref{sec:shape}.  Equation~\eqref{eq:kernel_one} contains a double integration over scaled time variables $\uptau_1$ and $\uptau_2$. 
We can split the two time variables using the properties of trigonometric functions
\begin{subequations}
\begin{eqnarray}
&&    \cos(k\eta_-) \cos(p c_s\eta_-)\cos(\Tilde{p} c_s\eta_-) = \nonumber\\
&& \qquad\qquad\qquad = \frac{1}{4}\sum_{m, n = \pm 1} \cos(\omega_{mn}\eta_-) \label{eq:cos}\\
&& \qquad\qquad\qquad = \frac{1}{4}\sum_{m, n = \pm 1} \Big\{\cos(\omega_{m n}\eta_1) \cos(\omega_{m n}\eta_2) + \sin(\omega_{m n}\eta_1) \sin(\omega_{m n}\eta_2)\Big\}.\label{eq:coscoscos}
\end{eqnarray}
\end{subequations}
The expressions for the kernel functions are thereby reduced to the calculation of integrals in the form
\begin{subequations}\label{eq:cos_n}
\begin{eqnarray}
    \int_{\uptau_*}^{\uptau_{\text{end}}} d\uptau \, \frac{\cos(\omega_{m n}\uptau)}{\uptau^{1+\nu}} &=& - \left\vert \omega_{mn}\right\vert^\nu  \Big[\operatorname{ci}_{-\nu}(\omega_{mn}\uptau_{\text{end}}) - \operatorname{ci}_{-\nu}(\omega_{mn}\uptau_*) \Big], \\
    \int_{\uptau_*}^{\uptau_{\text{end}}} d\uptau \, \frac{\sin(\omega_{m n}\uptau)}{\uptau^{1+\nu}} &=&  - \left\vert \omega_{mn}\right\vert^\nu \Big[\operatorname{si}_{-\nu}(\omega_{mn}\uptau_{\text{end}}) - \operatorname{si}_{-\nu}(\omega_{mn}\uptau_*) \Big], 
\end{eqnarray} 
\end{subequations}
with the generalized trigonometric integral functions defined by equation~\eqref{eqs:trigo}. Finally, this gives the kernel 
\begin{equation}\label{eq:kernel_app}
    \Delta (z, x, \Tilde{x}, \uptau_*, \uptau_{\text{end}}) = \frac{1}{8} \sum_{m, n = \pm 1} \left\vert\omega_{mn} \uptau_*\right\vert^{2\nu} \left[\left(\operatorname{ci}_{-\nu}(\omega_{m n} \uptau)\Big\vert^{\uptau_{\text{end}}}_{\uptau_*} \right)^2  + \left( \operatorname{si}_{-\nu}(\omega_{m n} \uptau)\Big\vert^{\uptau_{\text{end}}}_{\uptau_*} \right)^2 \right]
\end{equation}
as in equation~\eqref{eqs:kernel_omega}.

\paragraph{Approximation at low frequency.}
In the limit $k/p \rightarrow 0$ we have $\bm{q} = \bm{p} - \bm{k} \rightarrow \bm{p}$ and the sound wave momenta align. This way the kernel~\eqref{eq:kernel_one} simplifies to
\begin{equation}
    \Delta(z/x\rightarrow 0, x, \uptau_*, \uptau_{\text{end}}) = \frac{1}{4} \iint_{\uptau_*}^{\uptau_{\text{end}}}  \frac{d\uptau_1 d\uptau_2}{\uptau_*^2}  \left(\frac{\uptau_*^2}{\uptau_1 \uptau_2}\right)^{1+\nu}  \Big[ 1 + \cos(2c_s z\uptau_-) \Big].
\end{equation}
At this point, the properties of trigonometric functions allow the integration variables $\uptau_1$ and $\uptau_2$ to be separated and the integration to be carried out analytically, leading to equation~\eqref{eq:delta_k0}
\begin{eqnarray}
    \Delta (z/x\rightarrow 0, x, \uptau_*, \uptau_{\text{end}}) &=&  \frac{1}{4\nu^2}\left[1 - \left(\frac{\uptau_*}{\uptau_{\text{end}}}\right)^\nu\right]^2 + \nonumber\\
    && + \frac{(2c_sx\uptau_*)^{2\nu}}{4} \left[\left(\operatorname{ci}_{-\nu}(2c_sx\uptau) \Big\vert^{\uptau_{\text{end}}}_{\uptau_*} \right)^2  + \left( \operatorname{si}_{-\nu}(2c_s x \uptau)\Big\vert^{\uptau_{\text{end}}}_{\uptau_*} \right)^2 \right].\qquad\label{eq:kernel_zero}
\end{eqnarray}

\paragraph{Approximation at intermediate frequency.}
Ref.~\cite{Sharma:2023mao} pointed out that the power spectrum of gravitational wave in the regime $1 \ll k\eta_* \ll k_p\eta_*$ is dominated by the odd terms in the kernel~\eqref{eq:kernel_app} where $\omega_{mn} = \omega_\pm \equiv z \pm c_s(x-\tilde{x})$. When $k\eta_* \gg 1$, the arguments of the generalized trigonometric integrals have typically very large values, so that we are motivated to consider the expansion 
\begin{subequations}\label{eq:trigo_large}
\begin{eqnarray}
    \operatorname{si}_{-\nu} (x) & \underset{x\gg 1}{=} & \frac{\cos(x)}{x^{1+\nu}} +(1+\nu)\frac{\sin(x)}{x^{2+\nu}} + o\left(\frac{1}{x^{3+\nu}}\right), \\
    \operatorname{ci}_{-\nu} (x) &\underset{x\gg 1}{=}& - \frac{\sin(x)}{x^{1+\nu}}  +(1+\nu) \frac{\cos(x)}{x^{2+\nu}} + o\left(\frac{1}{x^{3+\nu}}\right).
\end{eqnarray}
\end{subequations}
At leading order in $1/k\eta_*$ the kernel~\eqref{eq:kernel_app} approximates to
\begin{eqnarray}
    \Delta(z,x,\tilde{x},\uptau_*, \uptau_{\text{end}}) & \underset{k\eta_* \gg 1}{\simeq}& \frac{1}{8} \sum_{\pm}\left(\omega_{\pm} \uptau_*\right)^{-2} \Bigg[1 + \left(1+\frac{\detav}{\eta_*}\right)^{-2(1+\nu)} - \nonumber\\
    &&\qquad\qquad\qquad  -2 \left(1+\frac{\detav}{\eta_*}\right)^{-(1+\nu)} \cos\left[\omega_{\pm}(\uptau_{\text{end}} - \uptau_*)\right]\Bigg],
\end{eqnarray}
where we defined $\detav = \eta_{\text{end}} - \eta_*$.
The sound wave scaled momenta $x$ and $\tilde{x}$ provide the most contribution to the power spectrum~\eqref{eq:pgw} at values around the peak scale $z_\star = k_p R_* = 2\pi$. In the gravitational wave frequency domain where $k\ll k_p$, i.e. $z\ll 1$, we can approximate $\tilde{x} \simeq x-\mu z$, with $\mu = \hat{\bm{x}}\cdot\hat{\bm{z}}$, so that $\omega_\pm \simeq z(1\pm c_s\mu$) and the kernel 
\begin{eqnarray}\label{eq:kernel_prior}
    \Delta_{\text{int}}(z,\mu,\uptau_*, \uptau_{\text{end}}) & \underset{1\ll k\eta_* \ll k_p\eta_*}{\simeq}& \frac{1}{8} \sum_{\pm}\left(k\eta_*(1\pm \mu c_s)\right)^{-2} \Bigg[ 1 + \left(1+\frac{\detav}{\eta_*}\right)^{-2(1+\nu)} - \qquad\nonumber\\
    &&\qquad\qquad  -2 \left(1+\frac{\detav}{\eta_*}\right)^{-(1+\nu)} \cos\left[k\detav (1\pm \mu c_s)\right]\Bigg].
\end{eqnarray}
The behavior of the kernel~\eqref{eq:kernel_prior} strongly depends on the time duration of the source. We consider three different scenarios:
\begin{itemize}
    \item \textbf{Short-lasting source ($\bm{\detav \ll \eta_*, \; k_p\detav \sim 1}$):} For short source duration, in the regime $1\ll k\eta_*\ll k_p\eta_*$,  we can perform a new Taylor expansion in the small parameter $\detav/\eta_*$. At leading order we obtain
    \begin{equation}\label{eq:k_short}
        \Delta_{\text{int}}^{\text{short}}(z, \mu, \uptau_*, \uptau_{\text{end}})  \simeq \frac{1}{8} \sum_{\pm} \left[1  + \left(\frac{1+\nu}{k\eta_* (1\pm \mu c_s)}\right)^2\right] \left(\frac{\detav}{\eta_*}\right)^2 \; \underset{k\eta_* \gg 1}{\sim} \; \frac{1}{4}\left(\frac{\detav}{\eta_*}\right)^2.
    \end{equation}
    We notice that this result is consistent with equation~\eqref{eq:kernel_zero} when $\uptau_*/\uptau_{\text{end}}\rightarrow 0$.  

    \item  \textbf{Medium-lasting source ($\bm{\detav \ll \eta_*, \; k_p \detav \gg 1}$):} 
    This is the case when the sound waves oscillate many times during the acoustic phase, and gravitational wave mode in the range $1\ll k\eta_*\ll k_p\eta_*$ oscillate with a period roughly of the order of the source duration. The oscillatory behavior of the periodic function becomes now important and,  at linear order in $\detav/\eta_*$, we find
    \begin{equation}\label{eq:k_mod}
        \Delta_{\text{int}}^{\text{medium}}(z,\mu, \uptau_*, \uptau_{\text{end}})  \simeq \frac{1}{2} \Big[1- (1+\nu)\frac{\detav}{\eta_*}  \Big] \sum_{\pm}    \left(\frac{\sin\left(k\detav(1\pm \mu c_s)/2\right)}{k\eta_* (1\pm \mu c_s)}\right)^2.
    \end{equation}
The oscillatory behavior of the kernel~\eqref{eq:k_mod} is relevant when $k\detav \sim \mathcal{O}(1)$. As we increase the number of gravitational wave oscillations within the time the source is active, which can be achieved increasing either $\detav$ or $k$, the oscillatory behavior of the kernel becomes less and less important for the power spectrum~\eqref{eq:pgw_1}, which becomes only sensitive to its average $\sin^2(x) \rightarrow 1/2$ when $x\gg 1$.

    \item  \textbf{Long-lasting source ($\bm{\detav \gg \eta_*}$):} When the source lasts much longer than a Hubble time, all the contributions proportional to $\eta_*/\eta_{\text{end}}$ in the expression~\eqref{eq:kernel_prior} can be neglected, so that
\begin{equation}\label{eq:ker_long}
    \Delta_{\text{int}}^{\text{long}}(z,\mu, \uptau_*, \uptau_{\text{end}})  \simeq \frac{1}{8} \sum_{\pm}\left(k\eta_*(1\pm \mu c_s)\right)^{-2} = \frac{1}{4(k\eta_*)^2} \frac{1+c_s^2\mu^2}{(1-c_s^2\mu^2)^2}.
\end{equation}
\end{itemize}
Regardless of the duration of the source, the kernel~\eqref{eq:kernel_prior} depends on the sound wave momenta $x$ and $\tilde{x}$, only through the angle $\mu$. It is therefore convenient to rewrite the dimensionless spectral density~\eqref{eq:pgw} as
\begin{equation}
    \tilde{P}_{\text{gw}} (kR_*) \simeq \frac{\uptau_*}{\pi^2} \int_0^\infty dx \, \tilde{P}_v(x) 
    \int_{-1}^{1} d\mu (1-\mu^2)^2 \frac{x^4}{\tilde{x}^2} \tilde{P}_v(\tilde{x}) \Delta_{\text{int}} (z, \mu, \uptau_*, \uptau_{\text{end}}),
\end{equation}
where we used the fact that $\rho(z, x, \tilde{x}) = (1-\mu^2)^2 x^3 z^2 /\tilde{x}$ and $\tilde{x}\,d\tilde{x} = - xz\,d\mu$.
Since in this range of frequency $\tilde{x} \simeq x-z\mu$, at leading order in $z/x \ll 1$ we can separate the integration variables and approximate
\begin{equation}\label{eq:P_interm}
    \tilde{P}_{\text{gw}} (kR_*) \simeq 2\uptau_*\mathcal{I}_v \int_{-1}^{1} d\mu \, (1-\mu^2)^2  \Delta_{\text{int}} (z, \mu, \uptau_*, \uptau_{\text{end}}),
\end{equation}
with
\begin{equation}
    \mathcal{I}_v \equiv \frac{1}{2\pi^2} \int_0^\infty dx \, x^2 \tilde{P}^2_v(x). 
\end{equation}
Equation~\eqref{eq:P_interm} is a generic approximation that holds for every stationary source in the intermediate frequency range $1\ll k\eta_* \ll k_p\eta_*$. The specification of the source only affects the value of $\mathcal{I}_v$; with the particular choice of 
the fluid velocity spectral density in equation~\eqref{eq:Pv}, we obtain $\mathcal{I}_v = 1/32\pi^2$.

Remembering the expression of the gravitational wave power spectrum~\eqref{eq:pgw_1}, we can infer the power-law indices in the frequency range $1\ll k\eta_* \ll k_p\eta_*$. Since the kernel~\eqref{eq:k_short} does not depend on the gravitational wave wavenumber $k$, we expect a scaling $\mathcal{P}^{\text{short}}_{\text{gw}} \propto k^3$ when the source lasts less than a gravitational wave oscillation period. When the source duration becomes of the order of the gravitational wave oscillation period, the kernel~\eqref{eq:k_mod} becomes sensitive to the gravitational wave oscillations, and the power spectrum $\mathcal{P}^{\text{medium}}_{\text{gw}} \propto k \sin^2(k\eta_v/2)$, which also shows the appearance of a shallow slope. When the source lasts for several gravitational wave oscillations, the  spectrum becomes only sensitive to the average of the oscillations and, from equation~\eqref{eq:ker_long}, $\mathcal{P}^{\text{long}}_{\text{gw}} \propto k$.

Having separated the integration variables as in equation~\eqref{eq:P_interm}, it is possible to perform the integration over the polar angle $\mu$ analytically. For long-lasting source, for example, we find
\begin{equation}
    \tilde{P}_{\text{gw}} (kR_*) \simeq \frac{4}{3c_s^4}  \left[ 3-2c_s^2 - \frac{3}{2c_s}(1-c_s^2) \ln\left( \frac{1+c_s}{1-c_s}\right) \right]   \frac{\mathcal{I}_v}{\uptau_* z^2}.
\end{equation}

\paragraph{Approximation at high frequency}
In the regime $k \gtrsim k_p$, assuming that the source of shear stress remains approximately stationary for many Hubble times $\eta_{\text{end}}\gg \eta_*$, we can simplify the expression of the kernel~\eqref{eq:kernel_one} taking advantage of the oscillatory behavior of the trigonometric functions. Let us first perform a change integration variables $\uptau_+ = (\uptau_1 +\uptau_2)/2$ and $\uptau_- = \uptau_1-\uptau_2$, so that
\begin{equation}\label{eq:delta_hf}
    \Delta_{\text{high}}(z,x,\tilde{x},\uptau_*, \uptau_{\text{end}}) = \frac{\uptau_*^{2\nu}}{2} \int_{\uptau_*}^{\uptau_{end}} d\uptau_+ \int_{-\bar{\uptau}_+}^{\bar{\uptau}_+} d\uptau_-     \frac{\cos(z\uptau_-)\cos(c_s x\uptau_-)\cos(c_s \tilde{x}\uptau_-)}{\left(\uptau_+^2 - \uptau_-^2/4\right)^{1+\nu}},
\end{equation}
and the domain of integration is set by $\bar{\uptau}_+ = 2(\uptau_+ -\uptau_*)$. In Ref~\cite{Guo:2020grp}, it was shown that the sound waves decorrelate rapidly when $\uptau_- \gtrsim \uptau_*$, so that the fluid kinetic energy is transferred into gravitational waves mostly when $\uptau_- \ll \uptau_+$. We can then perform a Taylor expansion of the integrand in equation~\eqref{eq:delta_hf} and carry out the integration over $\uptau_+$ neglecting contributions of order $(\uptau_-/\uptau_+)^2$. Using further the relation~\eqref{eq:cos} to simplify the product of trigonometric functions, we get
\begin{equation}\label{eq:D}
    \Delta_{\text{high}}(z,x,\tilde{x},\uptau_*, \uptau_{\text{end}}) \simeq \frac{\uptau_*^{-1}}{8(1+2\nu)} \left[1-\left(\frac{\eta_*}{\eta_{\text{end}}}\right)^{1+2\nu} \right] \sum_{m,n = \pm 1}\int_{-\uptau_{\text{end}}}^{\uptau_{\text{end}}} d\uptau_- \cos(\omega_{mn} \uptau_-).
\end{equation}
For long-lasting sources $\eta_{\text{end}}\gg \eta_*$,  the integral over $\uptau_-$ approximates to a Dirac delta  function centered at $\omega_{mn}$. Since only the case $m = n = -1$ can realize $z +c_s(m x + n\Tilde{x}) = 0$, we have
\begin{equation}\label{eq:D_long}
    \Delta_{\text{high}}(z,x,\tilde{x},\uptau_*, \uptau_{\text{end}}) \simeq \frac{\pi}{4(1+2\nu)}\, \uptau_*^{-1} \left[1-\left(\frac{\eta_*}{\eta_{\text{end}}}\right)^{1+2\nu} \right] \delta\big(z -c_s(x+\tilde{x})\big).
\end{equation}
Let us consider now the dimensionless spectral density function~\eqref{eq:pgw}
\begin{equation}
    \tilde{P}_{\text{gw}} (kR_*) = \frac{\uptau_*}{\pi^2 z^3}  \int_0^\infty dx \int_{\vert x - z\vert}^{x+z}d\tilde{x} \, \rho(z, x, \tilde{x}) \tilde{P}_v(x) \tilde{P}_v(\tilde{x}) \Delta_{\text{high}} \left(z, x, \tilde{x}, \tau, \tau_*, \tau_{\text{end}}\right)
\end{equation}
The delta function in the kernel~\eqref{eq:D_long} allows us to perform the integration over the sound wave scaled wavenumber $\Tilde{x}$ and to write
\begin{equation}\label{eq:P_p}
    \tilde{P}_{\text{gw}} (kR_*) = \frac{1}{4\pi c_s z^3} \frac{1}{(1+2\nu)} \left[1-\left(\frac{\eta_*}{\eta_{\text{end}}}\right)^{1+2\nu} \right] \int_{x_-}^{x_+} dx \, \rho(z, x) \tilde{P}_v(x) \tilde{P}_v(x_+ + x_- -x) .
\end{equation}
with $x_\pm = z(1\pm c_s)/(2c_s)$ and
\begin{equation}\label{eq:rho(z,x)}
    \rho(z, x) = z^2 \left(\frac{1-c_s^2}{c_s^2}\right)^2 \frac{(x-x_+)^2(x-x_-)^2}{x(x_+ + x_- - x)}.
\end{equation}
The extrema of integration in equation~\eqref{eq:P_p} are set by considering that the delta function enforces $x = z/c_s - \tilde{x}$ and that $x-z \leq \tilde{x} \leq x+z$.

\section{Detail on the numerical integration}\label{sec:append_numerics}
The greatest challenge in the numeric evaluation of the spectral density function~\eqref{eq:pgw} is the accurate computation of the kernel. This can be computed by integrating numerically equation~\eqref{eq:kernel_one} or by using the analytic result in equation~\eqref{eqs:kernel_omega}. The latter option reduces the dimensions of the integral to evaluate, and guarantees a better performance in terms of execution time. In this Appendix we elucidate how this computation can be implemented in a Python script with SciPy. The generalized trigonometric functions $\operatorname{ci}_{-\nu}(x)$ and $\operatorname{si}_{-\nu}(x)$ are only defined in the SciPy library for the case $\nu=0$. For the cases $\nu\neq 0$, we need instead to relate the generalized trigonometric functions to other known geometric functions.
Let us start introducing the generalized exponential integral~\cite[\href{https://dlmf.nist.gov/8.19.E3}{(8.19.3)}]{NIST:DLMF}
\begin{equation}
    E_\nu(ix) = \int_1^\infty \frac{e^{-ixt}}{t^\nu}dt,
\end{equation}
which is related to the generalized trigonometric functions by
\begin{equation}
    E_\nu(ix) = x^{\nu-1}\big[\operatorname{ci}_{1-\nu} (x) - i \operatorname{si}_{1-\nu} (x) \big], \qquad x\in \mathbb{R},
\end{equation}
and to the incomplete Gamma function by~\cite[\href{https://dlmf.nist.gov/8.19.E1}{(8.19.1)}]{NIST:DLMF}
\begin{equation}
    E_\nu(z) = z^{\nu -1}\Gamma(1-\nu, z), \qquad z\in \mathbb{C},
\end{equation}
where 
\begin{equation}
    \Gamma(a,z) = \int_z^\infty t^{a-1} e^{-t} dt.
\end{equation}
The incomplete Gamma function can be further related to the Krummer confluent hypergeometric function~\cite{abramowitz1948handbook}~\cite[\href{https://dlmf.nist.gov/8.5.E1}{(8.5.1)}]{NIST:DLMF}
\begin{equation}
    \Gamma(a,z) = \Gamma(a) - \frac{z^a}{a} \, _1F_1(a,a+1,-z),
\end{equation}
so that
\begin{equation}\label{eq:E_hyper}
    E_{1+\nu} (z)  = z^{\nu}\left[ \Gamma(-\nu) + \frac{z^{-\nu}}{\nu}\, _1F_1(-\nu,1-\nu,-z) \right].
\end{equation}
Finally we can write
\begin{subequations}\label{eqs:gen_trig}
    \begin{eqnarray}
        \operatorname{ci}_{-\nu}(x) &=& x^{-\nu} \operatorname{Re}\{E_{1+\nu}(ix)\} = \operatorname{Re}\left\{ i^{\nu}\left[ \Gamma(-\nu) + \frac{(ix)^{-\nu}}{\nu}\, _1F_1(-\nu,1-\nu,-ix) \right]\right\}, \qquad\\
        \operatorname{si}_{-\nu}(x) &=& -x^{-\nu} \operatorname{Im}\{E_{1+\nu}(ix)\} = -  \operatorname{Im}\left\{ i^{\nu}\left[ \Gamma(-\nu) + \frac{(ix)^{-\nu}}{\nu}\, _1F_1(-\nu,1-\nu,-ix) \right]\right\}.\qquad
    \end{eqnarray}
\end{subequations}
Equations~\eqref{eqs:gen_trig} are the expressions for the generalized trigonometric integrals that we use to evaluate the kernel~\eqref{eqs:kernel_omega}. 

A special care must be taken to the case $x = 0$, where the factor $(ix)^{-\nu}$ in equations~\eqref{eqs:gen_trig} can cause divergence problems in the numerical evaluation. In this case the generalized cosine integral can be evaluated analytically as
\begin{equation}
    \operatorname{ci}_{-\nu}(x\rightarrow 0) =  x^{-\nu} \int_1^\infty \frac{\cos(xt)}{t^{1+\nu}}dt   = \frac{x^{-\nu}}{\nu},
\end{equation}
while the generalized sine integral trivially vanishes. The apparent $x^{-\nu}$ divergence cancels in kernel~\eqref{eq:kernel_app}, where the terms with $\omega_{mn} = 0$ are then evaluated as 
\begin{equation}
    \Delta_{mn} (\omega_{mn}=0) = \frac{1}{8\nu^2} \left[1-\left(\frac{\uptau_*}{\uptau_{\text{end}}}\right)^{\nu} \right]^2 .
\end{equation}

\paragraph{Approximation for $\vert z\vert >20$:} the routine used by the SciPy library to compute the Krummer confluent hypergeometric function $_1F_1$ becomes unreliable at large arguments $\vert z\vert\geq 30$. In this regime we prefer to use instead the analytic approximation~\cite{abramowitz1948handbook} 
\begin{equation}
    _1F_1(a,b,z) \sim \Gamma(b) \left[\frac{e^z z^{a-b}}{\Gamma(a)} + \frac{(-z)^{-a}}{\Gamma(b-a)}\right]
\end{equation}
to evaluate the hypergeometric function in equation~\eqref{eq:E_hyper} as
\begin{equation}\label{approx}
    _1F_1(-\nu,1-\nu,-ix) = \Gamma(1-\nu) \left[\frac{e^{-ix} (-ix)^{-1}}{\Gamma(-\nu)} + \frac{(ix)^{\nu}}{\Gamma(1)}\right].
\end{equation}

\bibliographystyle{JHEP}
\bibliography{biblio}

\providecommand{\href}[2]{#2}\begingroup\raggedright\begin{thebibliography}{10}

\bibitem{Maggiore:1999vm}
M.~Maggiore, \emph{{Gravitational wave experiments and early universe cosmology}}, \href{https://doi.org/10.1016/S0370-1573(99)00102-7}{\emph{Phys. Rept.} {\bfseries 331} (2000) 283} [\href{https://arxiv.org/abs/gr-qc/9909001}{{\ttfamily gr-qc/9909001}}].

\bibitem{Caprini:2019egz}
C.~Caprini et~al., \emph{{Detecting gravitational waves from cosmological phase transitions with LISA: an update}}, \href{https://doi.org/10.1088/1475-7516/2020/03/024}{\emph{JCAP} {\bfseries 03} (2020) 024} [\href{https://arxiv.org/abs/1910.13125}{{\ttfamily 1910.13125}}].

\bibitem{LIGOScientific:2007fwp}
{\scshape LIGO Scientific} collaboration, \emph{{LIGO: The Laser interferometer gravitational-wave observatory}}, \href{https://doi.org/10.1088/0034-4885/72/7/076901}{\emph{Rept. Prog. Phys.} {\bfseries 72} (2009) 076901} [\href{https://arxiv.org/abs/0711.3041}{{\ttfamily 0711.3041}}].

\bibitem{Moore:2014lga}
C.J.~Moore, R.H.~Cole and C.P.L.~Berry, \emph{{Gravitational-wave sensitivity curves}}, \href{https://doi.org/10.1088/0264-9381/32/1/015014}{\emph{Class. Quant. Grav.} {\bfseries 32} (2015) 015014} [\href{https://arxiv.org/abs/1408.0740}{{\ttfamily 1408.0740}}].

\bibitem{Caprini:2024ofd}
C.~Caprini, O.~Pujol\`as, H.~Quelquejay-Leclere, F.~Rompineve and D.A.~Steer, \emph{{Primordial gravitational wave backgrounds from phase transitions with next generation ground based detectors}},  \href{https://arxiv.org/abs/2406.02359}{{\ttfamily 2406.02359}}.

\bibitem{EPTA:2023fyk}
{\scshape EPTA, InPTA:} collaboration, \emph{{The second data release from the European Pulsar Timing Array - III. Search for gravitational wave signals}}, \href{https://doi.org/10.1051/0004-6361/202346844}{\emph{Astron. Astrophys.} {\bfseries 678} (2023) A50} [\href{https://arxiv.org/abs/2306.16214}{{\ttfamily 2306.16214}}].

\bibitem{EPTA:2023xxk}
{\scshape EPTA} collaboration, \emph{{The second data release from the European Pulsar Timing Array: V. Implications for massive black holes, dark matter and the early Universe}},  \href{https://arxiv.org/abs/2306.16227}{{\ttfamily 2306.16227}}.

\bibitem{NANOGrav:2023gor}
{\scshape NANOGrav} collaboration, \emph{{The NANOGrav 15 yr Data Set: Evidence for a Gravitational-wave Background}}, \href{https://doi.org/10.3847/2041-8213/acdac6}{\emph{Astrophys. J. Lett.} {\bfseries 951} (2023) L8} [\href{https://arxiv.org/abs/2306.16213}{{\ttfamily 2306.16213}}].

\bibitem{NANOGrav:2023hvm}
{\scshape NANOGrav} collaboration, \emph{{The NANOGrav 15 yr Data Set: Search for Signals from New Physics}}, \href{https://doi.org/10.3847/2041-8213/acdc91}{\emph{Astrophys. J. Lett.} {\bfseries 951} (2023) L11} [\href{https://arxiv.org/abs/2306.16219}{{\ttfamily 2306.16219}}].

\bibitem{LISA:2017pwj}
{\scshape LISA} collaboration, \emph{{Laser Interferometer Space Antenna}},  \href{https://arxiv.org/abs/1702.00786}{{\ttfamily 1702.00786}}.

\bibitem{LISACosmologyWorkingGroup:2022jok}
{\scshape LISA Cosmology Working Group} collaboration, \emph{{Cosmology with the Laser Interferometer Space Antenna}},  \href{https://arxiv.org/abs/2204.05434}{{\ttfamily 2204.05434}}.

\bibitem{Caprini:2015zlo}
C.~Caprini et~al., \emph{{Science with the space-based interferometer eLISA. II: Gravitational waves from cosmological phase transitions}}, \href{https://doi.org/10.1088/1475-7516/2016/04/001}{\emph{JCAP} {\bfseries 04} (2016) 001} [\href{https://arxiv.org/abs/1512.06239}{{\ttfamily 1512.06239}}].

\bibitem{Kirzhnits:1972iw}
D.A.~Kirzhnits, \emph{{Weinberg model in the hot universe}}, {\emph{JETP Lett.} {\bfseries 15} (1972) 529}.

\bibitem{Kirzhnits:1972ut}
D.A.~Kirzhnits and A.D.~Linde, \emph{{Macroscopic Consequences of the Weinberg Model}}, \href{https://doi.org/10.1016/0370-2693(72)90109-8}{\emph{Phys. Lett. B} {\bfseries 42} (1972) 471}.

\bibitem{Kirzhnits:1976ts}
D.A.~Kirzhnits and A.D.~Linde, \emph{{Symmetry Behavior in Gauge Theories}}, \href{https://doi.org/10.1016/0003-4916(76)90279-7}{\emph{Annals Phys.} {\bfseries 101} (1976) 195}.

\bibitem{Kibble:1980mv}
T.W.B.~Kibble, \emph{{Some Implications of a Cosmological Phase Transition}}, \href{https://doi.org/10.1016/0370-1573(80)90091-5}{\emph{Phys. Rept.} {\bfseries 67} (1980) 183}.

\bibitem{Cohen:1990it}
A.G.~Cohen, D.B.~Kaplan and A.E.~Nelson, \emph{{Baryogenesis at the weak phase transition}}, \href{https://doi.org/10.1016/0550-3213(91)90395-E}{\emph{Nucl. Phys. B} {\bfseries 349} (1991) 727}.

\bibitem{Cline:2018fuq}
J.M.~Cline, \emph{{TASI Lectures on Early Universe Cosmology: Inflation, Baryogenesis and Dark Matter}}, {\emph{PoS} {\bfseries TASI2018} (2019) 001} [\href{https://arxiv.org/abs/1807.08749}{{\ttfamily 1807.08749}}].

\bibitem{Cline:2006ts}
J.M.~Cline, \emph{{Baryogenesis}},  in \emph{{Les Houches Summer School - Session 86: Particle Physics and Cosmology: The Fabric of Spacetime}}, 9, 2006 [\href{https://arxiv.org/abs/hep-ph/0609145}{{\ttfamily hep-ph/0609145}}].

\bibitem{Kodama:1982sf}
H.~Kodama, M.~Sasaki and K.~Sato, \emph{{Abundance of Primordial Holes Produced by Cosmological First Order Phase Transition}}, \href{https://doi.org/10.1143/PTP.68.1979}{\emph{Prog. Theor. Phys.} {\bfseries 68} (1982) 1979}.

\bibitem{Hawking:1982ga}
S.W.~Hawking, I.G.~Moss and J.M.~Stewart, \emph{{Bubble Collisions in the Very Early Universe}}, \href{https://doi.org/10.1103/PhysRevD.26.2681}{\emph{Phys. Rev. D} {\bfseries 26} (1982) 2681}.

\bibitem{Khlopov:1998nm}
M.Y.~Khlopov, R.V.~Konoplich, S.G.~Rubin and A.S.~Sakharov, \emph{{Formation of black holes in first order phase transitions}},  \href{https://arxiv.org/abs/hep-ph/9807343}{{\ttfamily hep-ph/9807343}}.

\bibitem{Franciolini:2021nvv}
G.~Franciolini, \emph{{Primordial Black Holes: from Theory to Gravitational Wave Observations}}, Ph.D. thesis, Geneva U., Dept. Theor. Phys., 2021.
\newblock \href{https://arxiv.org/abs/2110.06815}{{\ttfamily 2110.06815}}.
\newblock 10.13097/archive-ouverte/unige:156136.

\bibitem{Liu:2021svg}
J.~Liu, L.~Bian, R.-G.~Cai, Z.-K.~Guo and S.-J.~Wang, \emph{{Primordial black hole production during first-order phase transitions}}, \href{https://doi.org/10.1103/PhysRevD.105.L021303}{\emph{Phys. Rev. D} {\bfseries 105} (2022) L021303} [\href{https://arxiv.org/abs/2106.05637}{{\ttfamily 2106.05637}}].

\bibitem{Lewicki:2023ioy}
M.~Lewicki, P.~Toczek and V.~Vaskonen, \emph{{Primordial black holes from strong first-order phase transitions}},  \href{https://arxiv.org/abs/2305.04924}{{\ttfamily 2305.04924}}.

\bibitem{Hindmarsh:2020hop}
M.B.~Hindmarsh, M.~L\"uben, J.~Lumma and M.~Pauly, \emph{{Phase transitions in the early universe}}, \href{https://doi.org/10.21468/SciPostPhysLectNotes.24}{\emph{SciPost Phys. Lect. Notes} {\bfseries 24} (2021) 1} [\href{https://arxiv.org/abs/2008.09136}{{\ttfamily 2008.09136}}].

\bibitem{Witten:1980ez}
E.~Witten, \emph{{Cosmological Consequences of a Light Higgs Boson}}, \href{https://doi.org/10.1016/0550-3213(81)90182-6}{\emph{Nucl. Phys. B} {\bfseries 177} (1981) 477}.

\bibitem{Guth:1981uk}
A.H.~Guth and E.J.~Weinberg, \emph{{Cosmological Consequences of a First Order Phase Transition in the SU(5) Grand Unified Model}}, \href{https://doi.org/10.1103/PhysRevD.23.876}{\emph{Phys. Rev. D} {\bfseries 23} (1981) 876}.

\bibitem{Steinhardt:1980wx}
P.J.~Steinhardt, \emph{{The {Weinberg-Salam} Model and Early Cosmology}}, \href{https://doi.org/10.1016/0550-3213(81)90016-X}{\emph{Nucl. Phys. B} {\bfseries 179} (1981) 492}.

\bibitem{Steinhardt:1981ct}
P.J.~Steinhardt, \emph{{Relativistic Detonation Waves and Bubble Growth in False Vacuum Decay}}, \href{https://doi.org/10.1103/PhysRevD.25.2074}{\emph{Phys. Rev. D} {\bfseries 25} (1982) 2074}.

\bibitem{Witten:1984rs}
E.~Witten, \emph{{Cosmic Separation of Phases}}, \href{https://doi.org/10.1103/PhysRevD.30.272}{\emph{Phys. Rev. D} {\bfseries 30} (1984) 272}.

\bibitem{Hogan:1986qda}
C.J.~Hogan, \emph{{Gravitational radiation from cosmological phase transitions}}, {\emph{Mon. Not. Roy. Astron. Soc.} {\bfseries 218} (1986) 629}.

\bibitem{Kosowsky:1991ua}
A.~Kosowsky, M.S.~Turner and R.~Watkins, \emph{{Gravitational radiation from colliding vacuum bubbles}}, \href{https://doi.org/10.1103/PhysRevD.45.4514}{\emph{Phys. Rev. D} {\bfseries 45} (1992) 4514}.

\bibitem{Kosowsky:1992rz}
A.~Kosowsky, M.S.~Turner and R.~Watkins, \emph{{Gravitational waves from first order cosmological phase transitions}}, \href{https://doi.org/10.1103/PhysRevLett.69.2026}{\emph{Phys. Rev. Lett.} {\bfseries 69} (1992) 2026}.

\bibitem{Kosowsky:1992vn}
A.~Kosowsky and M.S.~Turner, \emph{{Gravitational radiation from colliding vacuum bubbles: envelope approximation to many bubble collisions}}, \href{https://doi.org/10.1103/PhysRevD.47.4372}{\emph{Phys. Rev. D} {\bfseries 47} (1993) 4372} [\href{https://arxiv.org/abs/astro-ph/9211004}{{\ttfamily astro-ph/9211004}}].

\bibitem{Huber:2008hg}
S.J.~Huber and T.~Konstandin, \emph{{Gravitational Wave Production by Collisions: More Bubbles}}, \href{https://doi.org/10.1088/1475-7516/2008/09/022}{\emph{JCAP} {\bfseries 09} (2008) 022} [\href{https://arxiv.org/abs/0806.1828}{{\ttfamily 0806.1828}}].

\bibitem{Jinno:2017fby}
R.~Jinno and M.~Takimoto, \emph{{Gravitational waves from bubble dynamics: Beyond the Envelope}}, \href{https://doi.org/10.1088/1475-7516/2019/01/060}{\emph{JCAP} {\bfseries 01} (2019) 060} [\href{https://arxiv.org/abs/1707.03111}{{\ttfamily 1707.03111}}].

\bibitem{Cutting:2018tjt}
D.~Cutting, M.~Hindmarsh and D.J.~Weir, \emph{{Gravitational waves from vacuum first-order phase transitions: from the envelope to the lattice}}, \href{https://doi.org/10.1103/PhysRevD.97.123513}{\emph{Phys. Rev. D} {\bfseries 97} (2018) 123513} [\href{https://arxiv.org/abs/1802.05712}{{\ttfamily 1802.05712}}].

\bibitem{Hindmarsh:2013xza}
M.~Hindmarsh, S.J.~Huber, K.~Rummukainen and D.J.~Weir, \emph{{Gravitational waves from the sound of a first order phase transition}}, \href{https://doi.org/10.1103/PhysRevLett.112.041301}{\emph{Phys. Rev. Lett.} {\bfseries 112} (2014) 041301} [\href{https://arxiv.org/abs/1304.2433}{{\ttfamily 1304.2433}}].

\bibitem{Hindmarsh:2015qta}
M.~Hindmarsh, S.J.~Huber, K.~Rummukainen and D.J.~Weir, \emph{{Numerical simulations of acoustically generated gravitational waves at a first order phase transition}}, \href{https://doi.org/10.1103/PhysRevD.92.123009}{\emph{Phys. Rev. D} {\bfseries 92} (2015) 123009} [\href{https://arxiv.org/abs/1504.03291}{{\ttfamily 1504.03291}}].

\bibitem{Hindmarsh:2016lnk}
M.~Hindmarsh, \emph{{Sound shell model for acoustic gravitational wave production at a first-order phase transition in the early Universe}}, \href{https://doi.org/10.1103/PhysRevLett.120.071301}{\emph{Phys. Rev. Lett.} {\bfseries 120} (2018) 071301} [\href{https://arxiv.org/abs/1608.04735}{{\ttfamily 1608.04735}}].

\bibitem{Hindmarsh:2017gnf}
M.~Hindmarsh, S.J.~Huber, K.~Rummukainen and D.J.~Weir, \emph{{Shape of the acoustic gravitational wave power spectrum from a first order phase transition}}, \href{https://doi.org/10.1103/PhysRevD.96.103520}{\emph{Phys. Rev. D} {\bfseries 96} (2017) 103520} [\href{https://arxiv.org/abs/1704.05871}{{\ttfamily 1704.05871}}].

\bibitem{Hindmarsh:2019phv}
M.~Hindmarsh and M.~Hijazi, \emph{{Gravitational waves from first order cosmological phase transitions in the Sound Shell Model}}, \href{https://doi.org/10.1088/1475-7516/2019/12/062}{\emph{JCAP} {\bfseries 12} (2019) 062} [\href{https://arxiv.org/abs/1909.10040}{{\ttfamily 1909.10040}}].

\bibitem{Jinno:2020eqg}
R.~Jinno, T.~Konstandin and H.~Rubira, \emph{{A hybrid simulation of gravitational wave production in first-order phase transitions}}, \href{https://doi.org/10.1088/1475-7516/2021/04/014}{\emph{JCAP} {\bfseries 04} (2021) 014} [\href{https://arxiv.org/abs/2010.00971}{{\ttfamily 2010.00971}}].

\bibitem{Jinno:2022mie}
R.~Jinno, T.~Konstandin, H.~Rubira and I.~Stomberg, \emph{{Higgsless simulations of cosmological phase transitions and gravitational waves}}, \href{https://doi.org/10.1088/1475-7516/2023/02/011}{\emph{JCAP} {\bfseries 02} (2023) 011} [\href{https://arxiv.org/abs/2209.04369}{{\ttfamily 2209.04369}}].

\bibitem{Kosowsky:2001xp}
A.~Kosowsky, A.~Mack and T.~Kahniashvili, \emph{{Gravitational radiation from cosmological turbulence}}, \href{https://doi.org/10.1103/PhysRevD.66.024030}{\emph{Phys. Rev. D} {\bfseries 66} (2002) 024030} [\href{https://arxiv.org/abs/astro-ph/0111483}{{\ttfamily astro-ph/0111483}}].

\bibitem{Gogoberidze:2007an}
G.~Gogoberidze, T.~Kahniashvili and A.~Kosowsky, \emph{{The Spectrum of Gravitational Radiation from Primordial Turbulence}}, \href{https://doi.org/10.1103/PhysRevD.76.083002}{\emph{Phys. Rev. D} {\bfseries 76} (2007) 083002} [\href{https://arxiv.org/abs/0705.1733}{{\ttfamily 0705.1733}}].

\bibitem{Caprini:2009fx}
C.~Caprini, R.~Durrer, T.~Konstandin and G.~Servant, \emph{{General Properties of the Gravitational Wave Spectrum from Phase Transitions}}, \href{https://doi.org/10.1103/PhysRevD.79.083519}{\emph{Phys. Rev. D} {\bfseries 79} (2009) 083519} [\href{https://arxiv.org/abs/0901.1661}{{\ttfamily 0901.1661}}].

\bibitem{Caprini:2009yp}
C.~Caprini, R.~Durrer and G.~Servant, \emph{{The stochastic gravitational wave background from turbulence and magnetic fields generated by a first-order phase transition}}, \href{https://doi.org/10.1088/1475-7516/2009/12/024}{\emph{JCAP} {\bfseries 12} (2009) 024} [\href{https://arxiv.org/abs/0909.0622}{{\ttfamily 0909.0622}}].

\bibitem{RoperPol:2019wvy}
A.~Roper~Pol, S.~Mandal, A.~Brandenburg, T.~Kahniashvili and A.~Kosowsky, \emph{{Numerical simulations of gravitational waves from early-universe turbulence}}, \href{https://doi.org/10.1103/PhysRevD.102.083512}{\emph{Phys. Rev. D} {\bfseries 102} (2020) 083512} [\href{https://arxiv.org/abs/1903.08585}{{\ttfamily 1903.08585}}].

\bibitem{Auclair:2022jod}
P.~Auclair, C.~Caprini, D.~Cutting, M.~Hindmarsh, K.~Rummukainen, D.A.~Steer et~al., \emph{{Generation of gravitational waves from freely decaying turbulence}}, \href{https://doi.org/10.1088/1475-7516/2022/09/029}{\emph{JCAP} {\bfseries 09} (2022) 029} [\href{https://arxiv.org/abs/2205.02588}{{\ttfamily 2205.02588}}].

\bibitem{Cutting:2019zws}
D.~Cutting, M.~Hindmarsh and D.J.~Weir, \emph{{Vorticity, kinetic energy, and suppressed gravitational wave production in strong first order phase transitions}}, \href{https://doi.org/10.1103/PhysRevLett.125.021302}{\emph{Phys. Rev. Lett.} {\bfseries 125} (2020) 021302} [\href{https://arxiv.org/abs/1906.00480}{{\ttfamily 1906.00480}}].

\bibitem{Weir:2017wfa}
D.J.~Weir, \emph{{Gravitational waves from a first order electroweak phase transition: a brief review}}, \href{https://doi.org/10.1098/rsta.2017.0126}{\emph{Phil. Trans. Roy. Soc. Lond. A} {\bfseries 376} (2018) 20170126} [\href{https://arxiv.org/abs/1705.01783}{{\ttfamily 1705.01783}}].

\bibitem{RoperPol:2022iel}
A.~Roper~Pol, C.~Caprini, A.~Neronov and D.~Semikoz, \emph{{Gravitational wave signal from primordial magnetic fields in the Pulsar Timing Array frequency band}}, \href{https://doi.org/10.1103/PhysRevD.105.123502}{\emph{Phys. Rev. D} {\bfseries 105} (2022) 123502} [\href{https://arxiv.org/abs/2201.05630}{{\ttfamily 2201.05630}}].

\bibitem{Sharma:2023mao}
R.~Sharma, J.~Dahl, A.~Brandenburg and M.~Hindmarsh, \emph{{Shallow relic gravitational wave spectrum with acoustic peak}}, \href{https://doi.org/10.1088/1475-7516/2023/12/042}{\emph{JCAP} {\bfseries 12} (2023) 042} [\href{https://arxiv.org/abs/2308.12916}{{\ttfamily 2308.12916}}].

\bibitem{RoperPol:2023dzg}
A.~Roper~Pol, S.~Procacci and C.~Caprini, \emph{{Characterization of the gravitational wave spectrum from sound waves within the sound shell model}},  \href{https://arxiv.org/abs/2308.12943}{{\ttfamily 2308.12943}}.

\bibitem{Espinosa:2010hh}
J.R.~Espinosa, T.~Konstandin, J.M.~No and G.~Servant, \emph{{Energy Budget of Cosmological First-order Phase Transitions}}, \href{https://doi.org/10.1088/1475-7516/2010/06/028}{\emph{JCAP} {\bfseries 06} (2010) 028} [\href{https://arxiv.org/abs/1004.4187}{{\ttfamily 1004.4187}}].

\bibitem{Cai:2023guc}
R.-G.~Cai, S.-J.~Wang and Z.-Y.~Yuwen, \emph{{Hydrodynamic sound shell model}}, \href{https://doi.org/10.1103/PhysRevD.108.L021502}{\emph{Phys. Rev. D} {\bfseries 108} (2023) L021502} [\href{https://arxiv.org/abs/2305.00074}{{\ttfamily 2305.00074}}].

\bibitem{Guo:2020grp}
H.-K.~Guo, K.~Sinha, D.~Vagie and G.~White, \emph{{Phase Transitions in an Expanding Universe: Stochastic Gravitational Waves in Standard and Non-Standard Histories}}, \href{https://doi.org/10.1088/1475-7516/2021/01/001}{\emph{JCAP} {\bfseries 01} (2021) 001} [\href{https://arxiv.org/abs/2007.08537}{{\ttfamily 2007.08537}}].

\bibitem{Ares:2020lbt}
F.R.~Ares, M.~Hindmarsh, C.~Hoyos and N.~Jokela, \emph{{Gravitational waves from a holographic phase transition}}, \href{https://doi.org/10.1007/JHEP04(2021)100}{\emph{JHEP} {\bfseries 21} (2020) 100} [\href{https://arxiv.org/abs/2011.12878}{{\ttfamily 2011.12878}}].

\bibitem{Baumann:2007zm}
D.~Baumann, P.J.~Steinhardt, K.~Takahashi and K.~Ichiki, \emph{{Gravitational Wave Spectrum Induced by Primordial Scalar Perturbations}}, \href{https://doi.org/10.1103/PhysRevD.76.084019}{\emph{Phys. Rev. D} {\bfseries 76} (2007) 084019} [\href{https://arxiv.org/abs/hep-th/0703290}{{\ttfamily hep-th/0703290}}].

\bibitem{Durrer:2004fx}
R.~Durrer, \emph{{Cosmological perturbation theory}}, \href{https://doi.org/10.1007/978-3-540-31535-3_2}{\emph{Lect. Notes Phys.} {\bfseries 653} (2004) 31} [\href{https://arxiv.org/abs/astro-ph/0402129}{{\ttfamily astro-ph/0402129}}].

\bibitem{Brandenberger:2003vk}
R.H.~Brandenberger, \emph{{Lectures on the theory of cosmological perturbations}}, \href{https://doi.org/10.1007/978-3-540-40918-2_5}{\emph{Lect. Notes Phys.} {\bfseries 646} (2004) 127} [\href{https://arxiv.org/abs/hep-th/0306071}{{\ttfamily hep-th/0306071}}].

\bibitem{Maggiore:2007ulw}
M.~Maggiore, \emph{{Gravitational Waves. Vol. 1: Theory and Experiments}}, Oxford University Press (2007), \href{https://doi.org/10.1093/acprof:oso/9780198570745.001.0001}{10.1093/acprof:oso/9780198570745.001.0001}.

\bibitem{Isaacson:1968hbi}
R.A.~Isaacson, \emph{{Gravitational Radiation in the Limit of High Frequency. I. The Linear Approximation and Geometrical Optics}}, \href{https://doi.org/10.1103/PhysRev.166.1263}{\emph{Phys. Rev.} {\bfseries 166} (1968) 1263}.

\bibitem{Isaacson:1968zza}
R.A.~Isaacson, \emph{{Gravitational Radiation in the Limit of High Frequency. II. Nonlinear Terms and the Ef fective Stress Tensor}}, \href{https://doi.org/10.1103/PhysRev.166.1272}{\emph{Phys. Rev.} {\bfseries 166} (1968) 1272}.

\bibitem{Dahl:2021wyk}
J.~Dahl, M.~Hindmarsh, K.~Rummukainen and D.J.~Weir, \emph{{Decay of acoustic turbulence in two dimensions and implications for cosmological gravitational waves}}, \href{https://doi.org/10.1103/PhysRevD.106.063511}{\emph{Phys. Rev. D} {\bfseries 106} (2022) 063511} [\href{https://arxiv.org/abs/2112.12013}{{\ttfamily 2112.12013}}].

\bibitem{Dahl:2024eup}
J.~Dahl, M.~Hindmarsh, K.~Rummukainen and D.~Weir, \emph{{Primordial acoustic turbulence: three-dimensional simulations and gravitational wave predictions}},  \href{https://arxiv.org/abs/2407.05826}{{\ttfamily 2407.05826}}.

\bibitem{NIST:DLMF}
``{\it NIST Digital Library of Mathematical Functions}.'' \url{https://dlmf.nist.gov/}, Release 1.2.1 of 2024-06-15.

\bibitem{Durrer:2003ja}
R.~Durrer and C.~Caprini, \emph{{Primordial magnetic fields and causality}}, \href{https://doi.org/10.1088/1475-7516/2003/11/010}{\emph{JCAP} {\bfseries 11} (2003) 010} [\href{https://arxiv.org/abs/astro-ph/0305059}{{\ttfamily astro-ph/0305059}}].

\bibitem{Cai:2018teh}
R.-G.~Cai and S.-J.~Wang, \emph{{Energy budget of cosmological first-order phase transition in FLRW background}}, \href{https://doi.org/10.1007/s11433-018-9216-7}{\emph{Sci. China Phys. Mech. Astron.} {\bfseries 61} (2018) 080411} [\href{https://arxiv.org/abs/1803.03002}{{\ttfamily 1803.03002}}].

\bibitem{Giombi:2023jqq}
L.~Giombi and M.~Hindmarsh, \emph{{General relativistic bubble growth in cosmological phase transitions}},  \href{https://arxiv.org/abs/2307.12080}{{\ttfamily 2307.12080}}.

\bibitem{Jinno:2024nwb}
R.~Jinno and J.~Kume, \emph{{Gravitational effects on fluid dynamics in cosmological first-order phase transitions}},  \href{https://arxiv.org/abs/2408.10770}{{\ttfamily 2408.10770}}.

\bibitem{ARFKEN2013643}
G.B.~Arfken, H.J.~Weber and F.E.~Harris, \emph{Chapter 14 - bessel functions},  in \emph{Mathematical Methods for Physicists (Seventh Edition)}, G.B.~Arfken, H.J.~Weber and F.E.~Harris, eds., (Boston), pp.~643--713, Academic Press (2013), \href{https://doi.org/https://doi.org/10.1016/B978-0-12-384654-9.00014-1}{DOI}.

\bibitem{abramowitz1948handbook}
M.~Abramowitz and I.A.~Stegun, \emph{Handbook of mathematical functions with formulas, graphs, and mathematical tables}, vol.~55, US Government printing office (1948).

\end{thebibliography}\endgroup
\end{document}